\documentclass[a4paper,11pt]{article}
\pdfoutput=1 

\usepackage{jheppub} 

\usepackage[T1]{fontenc} 
\usepackage{dsfont}
\usepackage{caption}
\usepackage{subcaption}
\usepackage{amsmath}
\newsavebox{\imagebox}
\newcommand{\hmod}{H^{\text{mod}}}

\allowdisplaybreaks

\title{\boldmath Modular parallel transport of multiple intervals\\ in 1+1-dimensional free fermion theory}

\author[a]{Bowen Chen}
\author[a]{Bart{\l}omiej Czech}
\author[b,c,d,e]{Ling-Yan Hung}
\author[f,c]{Gabriel Wong}

\affiliation[a]{Institute for Advanced Study, Tsinghua University, Beijing 100084, China}
\affiliation[b]{State Key Laboratory of Surface Physics, Fudan University, Shanghai 200433, China}
\affiliation[c]{Department of Physics and Center for Field Theory and Particle Physics, \\ Fudan University, Shanghai 200433, China}
\affiliation[d]{Institute for Nanoelectronic Devices and Quantum Computing, \\ Fudan University, Shanghai 200433, China}
\affiliation[e]{Yau Mathematical Sciences Center, Tsinghua University, Beijing 100084, China}
\affiliation[f]{Harvard Center of Mathematical Sciences and Applications, \\ Harvard University, Cambridge, MA 02138, USA}

\emailAdd{chenbw95@gmail.com, bartlomiej.czech@gmail.com, lyhung@fudan.edu.cn, gabrielwon@gmail.com}

\abstract{Modular parallel transport is a generalization of Berry phases, applied to modular (entanglement) Hamiltonians. Here we initiate the study of modular parallel transport for disjoint field theory regions. We study modular parallel transport in the kinematic space of multi-interval regions in the vacuum of 1+1-dimensional free fermion theory---one of the few theories for which modular Hamiltonians on disjoint regions are known. We compute explicitly the generators of modular parallel transport, and explain why their relatively simple form follows from a half-sided modular inclusion. We also compute explicitly the curvature two-form of modular parallel transport. We contrast all calculations with the expected behavior of modular parallel transport in holographic theories, emphasizing the role of non-local terms that couple distinct intervals.}

\begin{document} 
\maketitle
\flushbottom

\section{Introduction}
In the last decade properties of quantum entanglement in field theory have attracted much scrutiny. The many motivations for this interest include, among others, the study of topological phases of matter, holographic duality, and quantum computing. This paper applies to a new theory a recently invented tool that characterizes field theoretic entanglement: modular parallel transport \cite{mpt}.

The idea of modular parallel transport is that varying the entanglement of a field theory subregion posits a transport problem, which is familiar from the study of Berry phases \cite{Berry:1984jv}. Whereas Berry studied varying dynamical Hamiltonians, modular parallel transport is concerned with varying modular or entanglement Hamiltonians. Whereas Berry's transport fixes the ambiguous $U(1)$ phase of a quantum state, modular parallel transport fixes a larger ambiguity valued in the commutant of the modular Hamiltonian. Modular parallel transport can be studied both for variations of the state on one local region, and for varying the region itself while working in one fixed global state. The latter instance---which is also the subject of the present paper---plays a significant role in holographic duality: its holonomies encapsulate intuitive features of the bulk geometry, including lengths of curves \cite{Czech:2017zfq} and integral transforms of bulk curvature \cite{mpt}.

This paper studies modular parallel transport in a non-holographic setting. We consider varying a multi-interval region drawn from the global vacuum of the 1+1-dimensional free fermion theory. This instance of modular parallel transport is interesting for several reasons. First, explicitly computable examples of modular parallel transport are few and far between because knowing the exact modular Hamiltonian---a notoriously difficult problem---is a prerequisite. The free fermion is one of the few quantum field theories for which vacuum modular Hamiltonians are known for arbitrary, disjoint regions \cite{Casini:2009vk, Arias:2018tmw}. Second, modular parallel transport is best understood in holographic theories, whose entanglement simplifies at leading order in a small parameter conventionally denoted $1/N$. (In 1+1 dimensions, that parameter can be identified with $1/c$, where $c$ is the central charge.) Leading order expressions in $1/N$ denote bulk geometric quantities, including areas of surfaces and diffeomorphisms that perturb them. On the other hand, recent work has given evidence that $1/N$ corrections are essential for understanding several key problems of gravity, including the black hole information paradox \cite{islands}. This motivates exploring modular parallel transport beyond leading order in $1/N$. The free fermion is far from holographic settings, but it has the advantage of being computable exactly, not in an approximation or expansion. 

To help the reader digest our results, we highlight two qualitative questions to bear in mind. The first is about the degree of non-locality in modular parallel transport. In holographic duality, modular parallel transport is dual to bulk diffeomorphisms, which map one Ryu-Takayanagi (RT) \cite{rtref} surface to another and act everywhere orthogonally to the initial surface \cite{mpt}. Therefore, at leading order in $1/N$, modular parallel transport of disjoint regions must decouple into a collection of smaller, independent modular transport problems---one per connected component of the bulk Ryu-Takayanagi surface. When we include $1/N$ corrections, this decoupling disappears because at this order the modular Hamiltonian is sensitive to the entanglement of bulk fields; the generators of modular parallel transport then become non-local. We have no prior data on such non-local effects in modular parallel transport. The free fermion is the first---albeit imperfect---laboratory in which to inspect them. Note that this problem requires the study of modular parallel transport of disjoint regions, which almost uniquely singles out the free fermion. 

The second pertinent question concerns the half-sided modular inclusion \cite{Wiesbrock:1992mg, Borchers:1999}. It is a technical condition, which associates the embedding of a smaller algebra of observables within a larger one to a special form of quantum entanglement. In 1+1-dimensional quantum field theory, for algebras of observables in causal diamonds that differ by a null deformation at one endpoint, this condition is obeyed by the entanglement of the global vacuum state. That fact, as well as its generalization to higher-dimensional conformal field theories, plays an important role in holographic duality: it explains the emergent Poincar{\'e} symmetry of the bulk geometry \cite{Casini:2017roe,scramblingmodes,Leutheusser:2021frk}. For our purposes, it is important that modular inclusion---when it applies---simplifies the computation of modular parallel transport into the form given in equation~(\ref{dlambdaz}). In this paper, we first compute modular parallel transport without the benefit of modular inclusion. After finding (\ref{dlambdaz}), we verify that modular inclusion does in fact apply to multi-interval regions of the free fermion. In the Discussion, we contrast these results with the 1+1-dimensional free boson theory, in which modular parallel transport is not determined by modular inclusion.

The organization of the paper is as follows. Section~\ref{sec:review} reviews modular parallel transport and related concepts, as well as the multi-interval modular Hamiltonian in the 1+1-dimensional free fermion. Section~\ref{commutatorandmodes} covers preliminaries for the calculation: it computes commutators of frequently encountered terms, and constructs modes of definite modular energy. Section~\ref{2intMPT} contains the main result: it computes the two-interval modular parallel transport generators in equations~(\ref{dahmodz}-\ref{dbhmodz}) and in (\ref{dlambdaz}). We also make some novel observations about the modular parallel transport of single intervals, which is universal for all 1+1-dimensional conformal field theories in the vacuum. Section~\ref{sec:limits} examines the limiting behavior of the modular parallel transport generators as two constituent intervals shrink or get far apart. In contrast to exact single-interval answers and to holographic intuitions, we find that non-local terms remain important in all non-trivial limits. Section~\ref{sec:curvature} computes the curvature two-form of the two-interval modular parallel transport. The body of the paper focuses for clarity on the modular transport of double intervals, but the case of multiple intervals behaves similarly and is captured exactly  by equation~(\ref{MPTN}). We situate our calculations in a broader context---including the multi-interval generalization---in Section~\ref{sec:discussion}. 

\section{Review}
\label{sec:review}

After reviewing the concept of modular parallel transport \cite{mpt}, we present the vacuum modular Hamiltonian of a two-interval region in the 1+1-dimensional free fermion theory \cite{Casini:2009vk, Arias:2018tmw}.  We emphasize the multi-local symmetry, which underlies the structure of that modular Hamiltonian \cite{Rehren:2012wa,Tedesco:2014eaz} and which plays an important role in our analysis.

\subsection{Modular parallel transport}
\label{sec:mptrev}

Consider a continuous family of regions $R(\lambda)$ in a conformal field theory (CFT). Take a fixed global state $|\Psi\rangle \in \mathcal{H}_{\rm CFT}$ and, for each $R(\lambda)$, compute the two-sided modular Hamiltonian $H^{\rm mod}(\lambda)$. If the reduced density operators $\rho_{R(\lambda)} = {\rm Tr}_{R(\lambda)^c} |\Psi\rangle \langle \Psi|$ (as well as $\rho_{R(\lambda)^c}$) are well-defined then the two-sided modular Hamiltonian is
\begin{equation}
	H^{\rm mod}(\lambda) = H^{\textrm{mod, one-sided}}_{R(\lambda)} \otimes \mathds{1}_{{R(\lambda)}^c} - \mathds{1}_{R(\lambda)} \otimes H^{\textrm{mod, one-sided}}_{{R(\lambda)}^c},
	\label{hmodschem}
\end{equation}
where $\rho_{R(\lambda)} = \exp (-H^{\textrm{mod, one-sided}}_{R(\lambda)})$ and likewise for $\rho_{R(\lambda)^c}$. In general, however, the one-sided objects are not well-defined operators and we must work with the two-sided $H^{\rm mod}(\lambda)$, which always exists. In what follows we omit the epithet `two-sided' to avoid cluttering the text. 

The one-parameter family of modular Hamiltonians $H^{\rm mod}(\lambda)$ naturally define a transport problem, which can be described as follows. If the Hamiltonians can be diagonalized then we write them in the form
\begin{equation}
    \hmod (\lambda) = U^{\dagger} (\lambda) D (\lambda) U (\lambda), 
    \label{hmodfamily}
\end{equation}
where $D(\lambda)$ is a diagonal matrix of modular energies. Differentiating with respect to $\lambda$ gives:
\begin{equation}
    \partial_{\lambda} \hmod = \left[ \partial_{\lambda} U^{\dagger} U , \hmod  \right] + U^{\dagger} \partial_{\lambda} D U 
\end{equation}
The second term commutes with the modular Hamiltonian; following \cite{mpt, zm1} we call such operators \emph{zero modes}. It then makes sense to look for operators $V(\lambda)$, which satisfy:
\begin{equation}
    \partial_{\lambda} \hmod (\lambda) = \left[ V(\lambda), \hmod(\lambda)  \right] + \textrm{zero modes}
    \label{mptbasic}
\end{equation}
One way to describe $V$ is as a Heisenberg-picture `Hamiltonian,' which transports observables from one region to another in the family $R(\lambda)$.  

Certainly $\partial_{\lambda} U^{\dagger} U$ is one such $V$, but so is every operator of the form $(\partial_{\lambda} U^{\dagger} U + $~zero modes). In other words, equation~(\ref{mptbasic})---as a condition for an unknown $V(\lambda)$---has an ambiguity parameterized by zero modes of $\hmod(\lambda)$. Modular parallel transport fixes this ambiguity and selects a unique solution by the condition:
\begin{align}
	\left[ V , \hmod \right] &= (1-P_{0}) \partial_{\lambda} \hmod \label{mptdef1} \\
	P_{0}[V] &= 0 \label{mptdef2}
\end{align}
Here $P_0[V]$ is a projector onto zero-modes of $\hmod$, which we assume exists; see \cite{mpt} for its explicit definition. In the setup~(\ref{hmodfamily}), the solution of (\ref{mptdef1}-\ref{mptdef2}) is:
\begin{equation}
V = (1-P_0) [\partial_{\lambda} U^{\dagger} U] 
\label{mptU}
\end{equation}

If the $\hmod$s in (\ref{mptdef1}-\ref{mptdef2}) were dynamical Hamiltonians over a finite-dimensional Hilbert space then these equations would describe the non-Abelian generalization \cite{Wilczek:1984dh} of the Berry phase \cite{Berry:1984jv}. Indeed, the zero modes by which candidate solutions $V(\lambda)$ differ from one another are directly analogous to phases of ground states, which the system visits due to a slowly changing dynamical Hamiltonian. For this reason, the holonomies of modular parallel transport have been called modular Berry phases. 

In holographic duality, modular Hamiltonians $H(\lambda)$ are known to generate an orthogonal boost symmetry \cite{jlms} in the neighborhood of an extremal surface called Ryu-Takayanagi surface \cite{rtref}. Therefore, equation~(\ref{mptdef1}) can be understood in the bulk as a surface-changing diffeomorphism. The modular parallel condition~(\ref{mptdef2}) identifies a unique such diffeomorphism, which has no longitudinal (surface-preserving) and no local little-group (orthogonal boost) component \cite{mpt}. In this way, modular parallel transport gives a boundary handle on bulk diffeomorphisms in holographic duality. Part of the exercise in this paper is to see how the free fermion---which is not a holographic theory---differs from holographic theories in its modular parallel transport.

\subsection{Free fermion modular Hamiltonians and multi-local symmetry}
\label{sec:ffrev}

We now specialize to the free fermion theory in $1+1$-dimensional flat space. The theory decomposes into two decoupled chiral sectors, which can be described using coordinates $z = x-t$ and $\bar{z} = x+t$. For simplicity we focus on the left-moving sector, in which the (chiral component of) the fermion satisfies the anti-commutation relation 
\begin{equation}
    \{ \psi(z) , \psi^{\dagger} (w) \} = \delta(z-w)
    \end{equation}
    
We are interested in the vacuum modular Hamiltonian for a two-interval region $I_{1} \cup I_{2} = (a_{1} , b_{1} ) \cup (a_{2} , b_{2})$ in $z$-space. It was found in \cite{Casini:2009vk,Arias:2018tmw} to take the following form:
\begin{align}
    \hmod &= 2 \pi i \int dz\, \Big( h_{T}(z) T(z) + h_{C}(z) \psi^{\dagger}(z) \psi(\tilde{z}) \Big)
    \label{hmodexplicit} \\
 T(z) & = \frac{1}{2} \left(\partial \psi^{\dagger}(z) \psi (z) - \psi^{\dagger}(z) \partial \psi (z)\right) 
 \label{tzz} \\
    h_{T}(z) &= \left(\partial_{z} \log \left[ - \tfrac{(z-a_{1})(z-a_{2})}{(z-b_{1})(z-b_{2})} \right] \right)^{-1}  
    \label{deftlocal} \\
    h_{C}(z) &= \frac{\partial_{z} \bar{z}}{z - \bar{z}}\, h_{T}(z) 
\end{align}
Equation~(\ref{tzz}) and all operators in this paper are normal-ordered. Modular Hamiltonian~(\ref{hmodexplicit}) involves a function $\tilde{z}(z)$, which is:
\begin{align}
    \tilde{z}(z) &= \frac{a_1 a_2 (z - b_1 - b_2) - b_1 b_2 (z - a_1 - a_2)}{(a_1 + a_2 - b_1 - b_2) z - a_1 a_2 + b_1 b_2}
    \label{defztilde}
\end{align}
It is a map, which exchanges intervals $I_1$ and $I_2$ point-wise. It is straightforward to confirm that $\tilde{z}(a_1) = a_2$ and $\tilde{z}(b_1) = b_2$ and, more generally, that $\tilde{z}(\tilde{z}(z)) = z$. 

The new feature of this modular Hamiltonian is that, in addition to the local stress tensor $T(z)$, it involves the non-local current $\psi^\dagger(z) \psi(\tilde{z})$. We will probe this non-locality using modular parallel transport. Loosely speaking, we may think of this non-locality as a small $N$ precursor of something that is familiar from holographic duality: that disjoint intervals can have a connected entanglement wedge. The free fermion is, of course, not holographic, so the connection between these facts is rather crude. Nevertheless, we will find this point of view useful in Section~\ref{sec:curvature}, when we discuss the curvature of modular parallel transport. 

The modular Hamiltonian for $n>2$ intervals was also found in \cite{Casini:2009vk,Arias:2018tmw}. Its schematic structure is the same as (\ref{hmodexplicit}). 

\paragraph{Multi-local symmetry}
One fact, which enables an explicit computation of Hamiltonian~(\ref{hmodexplicit}), is that the algebra of observables of the chiral fermion on $n>1$ intervals has a vacuum-preserving isomorphism with the observable algebra of $n$ species of fermion on a single interval \cite{Rehren:2012wa,Tedesco:2014eaz}. In other words, the fermions on distinct intervals act like replicas of one another in all correlation functions. References~\cite{Rehren:2012wa,Tedesco:2014eaz} refer to this fact as a \emph{multi-local symmetry} and we adopt the same terminology. It is easiest to take the fiducial interval hosting the $n$ species of fermions to be $I = (0, \infty)$. We will call the isomorphism in question $\beta_n$:
\begin{equation}
	\beta_n : \mathcal{A}^{\otimes n}(I) \mapsto \mathcal{A}(I_{1} \cup I_{2} \cup \ldots I_n) \label{betadef}
\end{equation}
Explicit calculations in this paper will involve $\beta_2$, which we denote simply $\beta$ to avoid clutter. 

The existence of $\beta$ is a special, delicate property of the fermion. Isomorphisms like (\ref{betadef}) can sometimes be found for other theories, but vacuum-preserving isomorphisms are rare \cite{Longo:2003wy, Longo:2009mn}. One benefit of isomorphism (\ref{betadef}) is that it allows us to intertwine modular flows:
\begin{equation}
\sigma^{t}_{\mathcal{A}(I_{1} \cup I_{2})} = \beta \circ \sigma^{t}_{\mathcal{A}^{\otimes 2}(I)} \circ \beta^{-1} 
	\label{ittflow}
\end{equation}
Here $\sigma^{t}_{\mathcal{A}} (\cdot )$ defines the flow $\exp (it \hmod) (\cdot ) \exp (-it \hmod)$ in algebra $\mathcal{A}$ generated by the vacuum modular Hamiltonian. Note that the vacuum-preserving property of $\beta$ is essential to make sense of (\ref{ittflow}). The intertwining relation makes it manageable to compute modular flow in algebra $\mathcal{A}(I_{1} \cup I_{2})$ explicitly. That is because $\sigma^{t}_{\mathcal{A}^{\otimes 2}(I)}$ is a simple boost, as is known from the famous Bisognano-Wichman theorem \cite{Bisognano:1975ih}.

To compute $\sigma^{t}_{\mathcal{A}(I_{1} \cup I_{2})}$ explicitly we need to know $\beta$. The geometric space, which supports the domain of $\beta$ (where $I = (0, \infty)$ lives) will be coordinatized by $X$. The first aspect of $\beta$ is a geometric map
\begin{equation}
	    X(z) = - \frac{(z-a_{1})(z-a_{2})}{(z-b_{1})(z-b_{2})}\,, \label{xzmapping}
\end{equation}
which is two-to-one. It sends both $I_1 = (a_1, b_1)$ and $I_2=(a_2, b_2)$ to $I = (0, \infty)$. 
It has two inverses $z_1$ and $z_2$, which send $I$ back to $I_1$ (respectively $I_2$):
\begin{equation}
    z_{1,2}(X) = \frac{ a_{1} + a_{2} + (b_{1} + b_{2})X \mp \sqrt{(a_{1} + a_{2} + b_{1}X + b_{2}X)^{2} - 4(X+1) (b_{1}b_{2}X+a_{1}a_{2}) }  }{2(X+1)}
    \label{defz12}
    \end{equation}
The pairing in equation~(\ref{defztilde}) is the pairing of $z_1$ and $z_2$, that is $z_1(X(z_2)) = \tilde{z}(z_2)$. 

$X$-space is inhabited by two species of fermion, $\Psi_{1}(X)$ and $\Psi_{2}(X)$. We use capital $\Psi_i(X)$ to distinguish the two fermions in $I$ from the one fermion $\psi(z)$ in algebra $\mathcal{A}(I_{1} \cup I_{2})$. The anti-commutation relations in $X$-space are:
\begin{equation}
	\left\{  \Psi_{i}(X) , \Psi^{\dagger}_{j} (Y)  \right\} = \delta_{ij}\, \delta(X-Y)
\end{equation}
The action of the $X$-space modular flow is:
\begin{equation}
\sigma^{t}_{\mathcal{A}^{\otimes 2}(I)}(\Psi_i(X)) = e^{-\pi t}\, \Psi_i(X e^{-2 \pi t})
\label{defboost}
\end{equation}

We are now ready to exhibit $\beta$ explicitly:
\begin{equation}
	\beta \left[ \Psi_{i}(X) \right] = \sum_{j=1,2} U_{ij} (X) \sqrt{ \partial_{X}  z_{j}(X)}\, 
	\psi ( z_{j} (X) ) 
    \label{finalbeta}
\end{equation}
The matrix $U_{ij}$ is a gauge transformation $U(X) = \exp ({ \epsilon\, \theta(X)}) \in O(2)$, where $\epsilon = \epsilon_{ij}$ is the anti-symmetric tensor and the gauge parameter reads:
\begin{equation}
    \theta(X) =
    \int^{X} d\tilde{X} \frac{\sqrt{ \partial_{X} z_{1}(\tilde{X})\, \partial_{X} z_{2}(\tilde{X})}}{z_{1}(\tilde{X}) - z_{2}(\tilde{X})}
    \label{deftheta}
\end{equation}
This isomorphism maps (\ref{hmodexplicit}) to:
\begin{equation}
    \beta^{-1} \circ \hmod (z) = \int dX X \Big(T_{11}(X) + T_{22}(X)\Big)
    \label{hmodximage}
\end{equation}
Here $T_{11}$ and $T_{22}$ are the local stress tensors of $\Psi_1(X)$ and $\Psi_2(X)$, which have the same functional form as equation~(\ref{tzz}). Operator~(\ref{hmodximage}) generates boosts in $X$-space.

The gauge transformation (\ref{deftheta}), which is part of $\beta$, can be understood geometrically. A simple push-forward of the boost action of $\sigma^{t}_{\mathcal{A}^{\otimes 2}(I)}$ to $z$-space is singular when $dX/dz = 0$. This happens at $z = z_\pm$, which equal:
\begin{align}
z_{\pm} & = \frac{b_1 b_2 - a_1 a_2 \pm i \sqrt{-\xi}}{b_1 + b_2 - a_1 - a_2} 
\label{defzpm}\\
{\rm where}~~~\xi &= -(b_1 - a_1) (b_1 - a_2) (b_2 - a_1) (b_2 - a_2) > 0 
    \label{defxi} 
\end{align}
Gauge transformation~(\ref{deftheta}) resolves this singularity by inducing non-trivial monodromies
\begin{equation}
    \Psi_{1} (e^{2\pi i} X_{\pm}) = \pm \Psi_{2} (X_{\pm})
    \qquad {\rm and} \qquad
    \Psi_{2} (e^{2\pi i} X_{\pm}) = \mp\Psi_{1} (X_{\pm}),
\end{equation}
around the points $X_\pm = X(z_\pm)$. For more details, see \cite{Wong:2018svs}.

\paragraph{Modular parallel transport for multiple intervals in the free fermion---a guess}
Equation~(\ref{ittflow}) resembles the structure considered in equation~(\ref{hmodfamily}). In particular, $D$ appears in the same way as the $X$-space modular Hamiltonian~(\ref{hmodximage}) whereas the isomorphism $\beta$ roughly plays the role of $U^\dagger(\lambda)$. We remind the reader that $\beta$ depends on $\lambda = (a_1, b_1, a_2, b_2)$ through the definition of $X(z)$ and $\theta(X)$. When the context requires it, we will emphasize this dependence by writing $\beta_\lambda$. 

This parallel suggests that the modular parallel transport operator in the free fermion might be simply:
\begin{equation}
    V = 
    \left( 1 - P_{0} \right) \left[ \partial_{\lambda} \beta  \beta^{-1} \hmod \right]
    \label{mptderivative}
\end{equation}
The validity of this guess is not automatic, however, because (\ref{mptU}) and (\ref{mptderivative}) are not exact analogues. Unlike the $U(\lambda)$ in (\ref{mptU}), $\beta_\lambda$ does not constitute a legal operator in any theory. Instead, it is a map from operators in one Hilbert space to operators in another:
\begin{equation}
    \beta^{-1}\left[ \psi(z_{i}) \right] = \sum_{j} (U^{-1})_{ij} (X) \,\sqrt{\partial_{z_{i}} X(z)  }\, 
    \Psi_{j} (X) \big|_{X=X(z_i)}
\label{defbetainverse}
\end{equation}
Even the expression $d\lambda\, \partial_{\lambda} \beta$ must be interpreted with care. We will take it to mean the difference between images under $\beta_{\lambda+d\lambda}$ and $\beta_{\lambda}$ for infinitesimal $d\lambda$. Note that this derivative is complicated: $\lambda$-dependence is everywhere, including the $SO(2)$ rotation $U_{ij}$, the conformal factor $\sqrt{\partial_{z_{i}} X }$ and the $X$-position of the fermion.

Second, the $X(z)$ in equation~(\ref{xzmapping}) is not unique. If we rescale $X$ by a function of interval endpoints then $\tilde{X} = f(a_i, b_i) X$ has all the same properties as the original $X$. Such a rescaling would define a new, equally valid isomorphism $\beta$, which could potentially change (\ref{mptderivative}). We will see in Section~\ref{sec:1int} that artefacts of such redefinitions of $\beta$ are killed off by the projector in (\ref{mptderivative}). 

\section{Commutators and modular modes} 
\label{commutatorandmodes}
In solving for modular parallel transport, we need to separate modular zero modes---operators that commute with the modular Hamiltonian---from all others. This section characterizes the eigenmodes, which we will encounter. We emphasize the simplifications, which occur in the uniformized $X$-coordinate.  

\subsection{Commutators}
\label{sec:commutators}
Because we work in the free theory, all commutators follow from the fermionic anticommutation relations
\begin{align}
    \left\{ \psi^{\dagger}(z) , \psi (w) \right\} & = \delta(z-w) \\
    \left\{ \Psi_{i}^{\dagger}(X) , \Psi_{j} (Y) \right\} & =  \delta_{ij}\, \delta(X-Y)
    \label{basicxcomm}
\end{align}
by applying the operator equality:
\begin{equation}
    \left[ AB , CD \right] = A \left\{ B,C \right\} D - AC \left\{ B,D \right\} + \left\{ A,C \right\} DB - C \left\{ A,D \right\} B
\end{equation}

\paragraph{$X$-coordinate commutators}
We focus on operators of the form:
\begin{equation}
    \int dX \sum_{ij} \left( A_{ij}(X) T_{ij}(X) + B_{ij}(X) \Psi_{i}^{\dagger}(X) \Psi_{j}(X)  \right)
    \label{ourterms}
\end{equation}
where:
\begin{equation}
T_{ij}(X) = \frac{1}{2} :\partial \Psi_{i}^{\dagger}(X) \Psi_{j} (X) - \Psi_{i}^{\dagger}(X) \partial \Psi_{j} (X):
\label{deftij}
\end{equation}
In principle, terms quartic and higher in the fermions, and with higher order derivatives, can appear. However, $\beta^{-1} \circ \partial_{\lambda} \hmod$ is of the form~(\ref{ourterms}) and analyzing that expression suffices for our purposes. Any commutator of terms featured in~(\ref{ourterms}) is a combination of the following commutators, which in turn are derived from~(\ref{basicxcomm}):
\begin{align}
     \left[ \int f \Psi_{i}^{\dagger} \Psi_{j} , \int g \Psi_{k}^{\dagger} \Psi_{l}\right] 
     &=  \int  fg \,\big( \delta_{kj}\, \Psi_{i}^{\dagger} \Psi_{l} - \Psi_{k}^{\dagger} \Psi_{j}\, \delta_{il} \big)  \label{com1} \\
     \left[ \int f T_{ij} , \int g \Psi_{k}^{\dagger} \Psi_{l}\right] 
     &=  \int fg\, \big( \delta_{kj} T_{il} - T_{kj} \delta_{il} \big) 
     - \int \frac{fg'}{2} \,\big( \delta_{kj}\, \Psi_{i}^{\dagger} \Psi_{l} + \Psi_{k}^{\dagger} \Psi_{j}\, \delta_{il} \big) 
 \label{com2} \\
     \left[ \int f T_{ij} , \int g T_{kl}\right] 
     & = \int \frac{f'g - fg'}{2} \big( \delta_{kj} T_{il} + T_{kj} \delta_{il} \big) \nonumber \\
         &\,\,\,\,\,\,+ \int \left(\frac{f'' g  +f'g' + fg''}{4} - fg \right)
      \big(\delta_{kj}\, \psi^{\dagger}_{i} \Psi_{l} - \psi^{\dagger}_{k} \Psi_{j}\, \delta_{il} \big)  
     \label{com3}
\end{align}
All expressions in (\ref{com1}-\ref{com3}) are local in $X$. We omit the arguments and write $\int$ instead of $\int dX$ to lighten the notation.

\paragraph{$z$-coordinate commutators}
The counterpart of expression~(\ref{ourterms}) is:
\begin{equation}
    \int dz \left( a(z) T(z) + b(z) \tilde{T}(z) + c(z) \psi^{\dagger}(z) \psi(z) + d(z) \psi^{\dagger}(z) \psi(\tilde{z}) \right)
\end{equation}
There is a non-local fermion current $\psi^{\dagger}(z) \psi(\tilde{z})$ and a non-local stress tensor-like term $\tilde{T}$: 
\begin{align}
    \tilde{T}(z) = \frac{1}{2} \left( \partial \psi^{\dagger}(z) \psi(\tilde{z}) - \frac{\partial \tilde{z}}{\partial z} \psi^{\dagger}(z) \tilde{\partial} \psi(\tilde{z}) \right)
\end{align}
Because the $z$-picture fermions are non-locally coupled, we would now need as many as ten different commutators in place of (\ref{com1}-\ref{com3}), with many factors of $d\tilde{z} / dz$. We found the resulting expressions long and unilluminating.

\subsection{Modular modes} 
\label{XZMsubtraction}

\paragraph{Zero modes} A convenient notation for the zero modes of
\begin{equation}
\beta^{-1} \circ \hmod =   H_{1} + H_{2} = 2 \pi i \int dX X \big( T_{11}(X) + T_{22}(X) \big) 
\end{equation}
is to mark the basic fermion bilinear from which they originate in the subscript, and the number of derivatives in the superscript. In this convention, we have:
\begin{align}
Q^{(0)}_{ij} &= \int dX \Psi^{\dagger}_{i}(X) \Psi_{j}(X) \label{zerocurrent} \\
Q^{(1)}_{ij} & = \int dX X\, T_{ij}(X) \label{zerostress}
\end{align}
It is straightforward to confirm that (\ref{zerocurrent}-\ref{zerostress}) commute with $\hmod$ using~(\ref{com1}-\ref{com3}), for every choice of $i$ and $j$. It is possible to construct zero modes quartic and higher in the fermions, but these will suffice for our purposes. 

\paragraph{Scrambling modes}
We can use zero modes to construct so-called scrambling modes $G^{\pm}$. They are eigenmodes of the superoperator $[\hmod, \ldots]$, whose eigenvalues saturate the modular chaos bound \cite{scramblingmodes}. They satisfy the following relations:
\begin{equation}
    \left[ \beta^{-1} \circ \hmod , G^{\pm}   \right] = \pm 2 \pi i G^{\pm}
\end{equation}
In holography, scrambling modes are responsible for the emergence of Poincar{\'e} symmetry in the bulk \cite{Casini:2017roe,scramblingmodes,Leutheusser:2021frk}.

To get a $G^{-}$-type scrambling mode in our calculation, multiply the integrand in a zero mode by $X$. For a $G^{+}$-type scrambling mode, use $X^{-1}$ instead. This rule of thumb works because $\beta^{-1} \circ \hmod = 2\pi i L_0$ in $X$-space, so scrambling modes must be operators of conformal weight $\pm 1$. For example, we find:
\begin{align}
\left[ \beta^{-1} \circ \hmod, \int dX\, T_{ij} (X)   \right] 
&= + 2 \pi i \int dX\, T_{ij} (X) \label{scramb1} \\
\left[ \beta^{-1} \circ \hmod, \int dX X \Psi_i^{\dagger}(X) \Psi_j(X) \right] 
&=  - 2 \pi i \int dX X \Psi_i^{\dagger}(X) \Psi_j(X) \label{scramb2}
\end{align}
The other two possibilities are $\int dX X^2 T_{ij}$ and $\int dX X^{-1} \Psi^\dagger_i \Psi_j$; they too are scrambling modes. Scrambling modes play an important role in the remainder of this paper.

\section{Modular parallel transport on multiple intervals} 
\label{2intMPT}
We present an explicit solution for modular parallel transport of two disjoint intervals $I_{1} \cup I_{2} = (a_{1},b_{1}) \cup (a_{2},b_{2})$ in equation~(\ref{dlambdaz}). The answer ends up having the form~(\ref{mptderivative}), which we anticipated in Section~\ref{sec:ffrev}. To explain what this means---and how modular parallel transport accommodates the non-uniqueness of isomorphism $\beta$---we first revisit the single interval problem. 

%

\subsection{One-interval case revisited}
\label{sec:1int}
It is useful to revisit the single interval. The modular Hamiltonian for interval $(a,b)$ in the vacuum state of any two-dimensional conformal field theory is:
\begin{equation}
    \hmod_{\text{(1-int)}} = 2 \pi i \int dz \frac{(b-z)(z-a)}{b-a} T(z)
\end{equation}
Reference~\cite{mpt} computed the modular parallel transport in this case and found:
\begin{equation}
    V^{\text{(1-int)}}_{a} = - \int dz \frac{(b-z)^{2}}{(b-a)^{2}} T(z)
    \qquad {\rm and} \qquad
    V^{\text{(1-int)}}_{b} = - \int dz \frac{(z-a)^{2}}{(b-a)^{2}} T(z)
    \label{1intfermion}
\end{equation}
The subscripts refer to the kinematic space coordinate, which is being infinitesimally shifted. 
We can easily use the $z$-space commutators to verify equations~(\ref{1intfermion}). 

It is illuminating, however, to rephrase the problem in terms of:
\begin{equation}
X = - (z-a)/(z-b) \label{defx1int}
\end{equation}
Now denoting the map $z \to X$ as $\beta^{-1}$, we have:
\begin{equation}
 \beta^{-1} \circ \hmod_{\text{(1-int)}} = \frac{2 \pi i }{b-a} \int dX X\, T(X)
 \label{hmodx1int}
\end{equation}
The generators of modular parallel transport in the $X$-picture are:
\begin{equation}
\beta^{-1} \circ V^{\text{(1-int)}}_{a} = \frac{1}{b-a} \int dX T(X)
\qquad {\rm and} \qquad
\beta^{-1} \circ V^{\text{(1-int)}}_{b} = \frac{1}{b-a} \int dX X^2 T(X)
\label{mptx1int}
\end{equation}
These are just the scrambling modes, which we found in and below equation~(\ref{scramb1}). (Note that, when talking about a single interval, the indices $i,j$ play no role.)

Using the map $\beta^{-1}$ clarifies the $z$-space expressions for modular parallel transport given in equation~(\ref{1intfermion}). They are of the form $(\partial_\lambda \beta) \beta^{-1} \hmod_{\rm (1-int)}$, as anticipated in equation~(\ref{mptderivative}). To understand this assertion fully, consider modifying the definition of $X$ in equation~(\ref{defx1int}) by an additional, $(a,b)$-dependent rescaling:
\begin{equation}
\tilde{X} = - f(a,b)^{-1} (z-a) / (z-b)
\end{equation}
A logical notation for the map $z \to \tilde{X}$ is $\tilde{\beta}$. Because a global rescaling does not affect~(\ref{hmodx1int}), we have $\beta^{-1} \hmod = \tilde{\beta}^{-1} \circ \hmod$. If we now attempt to use equation~(\ref{mptderivative}) with $\tilde{\beta}^{-1}$ instead of $\beta^{-1}$, we will find:
\begin{equation}
    (\partial_{\lambda} \tilde{\beta}) \circ \tilde{\beta}^{-1} \circ \hmod 
    = (\partial_{\lambda} \beta) \circ \beta^{-1} \circ \hmod + \partial_{\lambda} f f^{-1} \hmod
\end{equation}
The two answers differ by a term, which is a modular zero mode---here, a multiple of the original modular Hamiltonian. The projector in equation~(\ref{mptderivative}) removes that zero mode so that modular parallel transport can be computed via equation~(\ref{mptderivative}) using both $\beta^{-1}$ and $\tilde{\beta}^{-1}$. What is interesting, however, is that there exists a global choice of $\beta^{-1}$---given in equation~(\ref{defx1int})---for which $(\partial_\lambda \beta) \beta^{-1} \hmod_{\rm (1-int)}$ is already zero mode-free. 

We will find that analogous statements continue to hold for multiple intervals:
\begin{itemize}
    \item Modular parallel transport is given by $(1-P^0) [(\partial_\lambda \beta) \beta^{-1} \hmod ]$, where $\beta^{-1}$ maps the one fermion on multiple intervals to multiple fermions on a single interval in $X$-space.
    \item When we take the standard uniformized coordinate $X = - \tfrac{\prod_{i} (z-a_{i})}{\prod_{i} (z-b_{i})}$ then modular parallel transport is given simply by $(\partial_{\lambda} \beta) \beta^{-1} \hmod$, without the need for zero mode subtraction.
        \item The generator of modular parallel transport, which shifts only left (respectively right) endpoints of intervals, is a scrambling mode with eigenvalue $- 2 \pi i$ (respectively $+ 2 \pi i$); see equation~(\ref{mptx1int}).  
\end{itemize}

\subsection{Two-interval modular parallel transport in $X$-space}

We map the two-interval modular Hamiltonian to $X$-space using $\beta^{-1}$, which was defined in equation~(\ref{finalbeta}). After this transformation, the modular Hamiltonian becomes:
\begin{equation}
\beta^{-1} \circ \hmod = 2 \pi i \int dX X \big( T_{1}(X) + T_{2}(X) \big) 
\end{equation}
Consider an operator of interest in $X$-space:
\begin{equation}
   O = \int dX \sum_{ij} \left( A_{ij}(X) T_{ij}(X) + B_{ij}(X) \Psi_{i}^{\dagger}(X) \Psi_{j}(X)  \right)
\end{equation}
Its commutator with the $X$-transformed modular Hamiltonian is:
\begin{align}
\left[ O , \beta^{-1} \circ \hmod \right] 
= 2\pi i \int dX \sum_{ij} \left[  \left(X A'_{ij}(X) -A_{ij}(X) \right)  T_{ij}(X) + X B'_{ij}(X) \Psi_{i}^{\dagger}(X) \Psi_{j}(X)  \right]
\label{ohcommx}
\end{align}
A nice feature of this equation is that the smearing functions do not mix. (This is in contrast to $z$-space, where no such decoupling happens.) Therefore, setting the commutator $[O, \beta^{-1} \circ \hmod]$ equal to the variation of $\beta^{-1} \hmod$ involves eight independent differential equations, which do not couple to one another. 

To simplify subsequent formulas, we introduce a shorthand notation:
\begin{equation}
\tau_{a_1} = \frac{(b_1 - a_2) (b_2 - a_2)}{(b_1 - a_1) (b_2 - a_1)}
\qquad {\rm and} \qquad
\tau_{b_1} = \frac{(b_2 - a_1) (b_2 - a_2)}{(b_1 - a_1) (b_1 - a_2)}
\end{equation}
The subscript indicates the repeated quantity in the denominator, so it is consistent notation to set $\tau_{a_2} = \tau_{a_1}^{-1}$ and $\tau_{b_2} = \tau_{b_1}^{-1}$. Using these expressions, the $a_1$-variation of the modular Hamiltonian for the two-interval region $(a_1, b_1) \cup (a_2, b_2)$ is:
\begin{align}
& \beta^{-1} \circ \partial_{a_1} \hmod \label{dahmodx} \\
& = \frac{2 \pi i}{b_2 - b_1} \int dX 
\bigg[ - \frac{b_2-a_2}{b_2-a_1}\, T_{11}(X) + \frac{b_1 - a_2}{b_1 - a_1}\, T_{22}(X) + \sqrt{-\tau_{a_1}}\Big( T_{12}(X) + T_{21}(X) \Big) \bigg] \nonumber
\end{align}

Comparing with equation~(\ref{scramb1}), we see that~(\ref{dahmodx}) is already a $G^-$-type scrambling mode and, in particular, that it contains no zero modes. Therefore, to let $O$ generate modular parallel transport, we set~(\ref{dahmodx}) equal to (\ref{ohcommx}). This leads to the following differential equations for $A_{ij}$:
\begin{align}
     X A'_{11}(X) -A_{11}(X) &= - \frac{b_2 - a_2}{(b_2 - b_1)(b_2 - a_1)} \label{diffeq1} \\
    X A'_{22}(X) -A_{22}(X) &= \frac{b_{1} - a_{2}}{(b_2 - b_1)(b_1 - a_1)} \\
    X A'_{12}(X) -A_{12}(X)&= \frac{\sqrt{-\tau_{a_1}}}{b_2 - b_1} =  X A'_{21}(X) -A_{21}(X) \label{diffeq4}
\end{align}
We also have $B_{ij}' = 0$. Even before solving these explicitly, we note that constants of integration enter the solutions in the form $A_{ij}(X) = A_{ij}^{(0)}(X) + a_{ij} X$ and $B_{ij}(X) = b_{ij}$. The freedom to set the constants of integration $a_{ij}$ and $b_{ij}$ is the freedom of adding zero modes to operator $O$. Modular parallel transport calls for a zero mode-free solution, which corresponds to some particular choice of the integration constants. For $B_{ij}$, the zero mode-free choice is clearly $B_{ij}(X) = b_{ij} = 0$.

One solution of~(\ref{diffeq1}-\ref{diffeq4}) is simply: 
\begin{equation}
   O = - \frac{1}{2\pi i}\, \beta^{-1} \circ \partial_{a_{1}} \hmod
\end{equation}
We already noted that the right hand side is zero mode-free. Therefore, this $O$ is the generator of modular parallel transport. For a general deformation of $(a_1, b_1) \cup (a_2, b_2)$, we find:
\begin{equation}
\beta^{-1} \circ V^{\text{(2-int)}}_{\lambda} = \frac{1}{2\pi i}\, \beta^{-1} \circ \left[ - \frac{\partial a_1}{\partial \lambda} \partial_{a_1} \hmod + \frac{\partial b_1}{\partial \lambda} \partial_{b_1} \hmod - \frac{\partial a_2}{\partial \lambda} \partial_{a_2} \hmod + \frac{\partial b_2}{\partial \lambda} \partial_{b_{2}} \hmod \right]
\label{dlambdax}
\end{equation}
Shifting left endpoints is generated by $G^-$-type scrambling modes whereas shifting right endpoints is generated by $G^+$-type scrambling modes. This is the third and final point, which was highlighted in Section~\ref{sec:1int}.

For completeness, we write down explicitly the operators, which generate modular parallel transport for an arbitrary variation of two intervals $(a_1, b_1) \cup (a_2, b_2)$. We list the explicit $X$-space smearing functions in the order $\{A_{11}, A_{12}, A_{21}, A_{22}\}$:
\begin{align}
\beta^{-1} \circ V_{a_{1}} 
&\supset \left\{ 
-\tfrac{b_2 - a_2}{(b_2 - b_1)(b_2 - a_1)},  
\tfrac{\sqrt{-\tau_{a_1}}}{b_2 - b_1}, 
\tfrac{\sqrt{-\tau_{a_1}}}{b_2 - b_1}, 
-\tfrac{b_{1} - a_{2}}{(b_1 - b_2)(b_1 - a_1)} \right\} \\
\beta^{-1} \circ V_{b_{1}} 
&\supset \left\{  0,0,0, 
\tfrac{(b_1 - b_2) X^{2} }{(b_1 - a_1) (b_1 - a_2) } \right\} \\
  \beta^{-1} \circ  V_{a_{2}} 
  &\supset \left\{ 
-\tfrac{b_2 - a_1}{ (b_2 - a_2) (b_2 - b_1)  } , 
\tfrac{\sqrt{-\tau_{a_2}}}{b_1 - b_2}, 
\tfrac{\sqrt{-\tau_{a_2}}}{b_1 - b_2}, 
-\tfrac{b_1 - a_1}{ (b_{1} - b_{2})(b_{1} - a_{2})  }  \right\} \\
\beta^{-1} \circ V_{b_{2}} 
&\supset \left\{ 
\tfrac{(b_2 - b_1) X^{2} }{(b_2 - a_1) (b_2 - a_2) }, 0,0,0  \right\} 
\end{align}

The three zeroes in $\beta^{-1} \circ V_{b_{1}}$ and $\beta^{-1} \circ V_{b_{2}}$ make it seem like left and right endpoints are on an entirely different footing. If true, this would be embarrassing because deforming $(a_1, b_1) \cup (a_2, b_2)$ should be equivalent to deforming their complement on the $X$-axis, which is also conformally equivalent to two intervals. Left endpoints of $(a_1, b_1) \cup (a_2, b_2)$ are right endpoints for the complement, and vice versa.

We will see that this asymmetry is an artifact of working in $X$-space. In equation~(\ref{xzmapping}), interchanging left and right endpoints corresponds to $X \to -X^{-1}$. When working with $Y = - 1/X$, the arrays $\{A_{11}, A_{12}, A_{21}, A_{22}\}$ take the analogous but distinct form, with the three zeros appearing in $\beta^{-1} \circ V_{a_{1}}$ and $\beta^{-1} \circ V_{a_{2}}$ instead. Both variations, however, take on a symmetric form when we work directly in $z$-space. We show this explicitly below.

\subsection{Two-interval modular parallel transport in $z$-space}
Here we inspect the variation of the modular Hamiltonian directly in $z$-space:
\begin{equation}
    \partial_{\lambda} \hmod = 2 \pi i \int dz \partial_{\lambda} \Big[h_{T}(z;\lambda) T(z) + \partial_{\lambda} h_{C}(z;\lambda) \psi^{\dagger}(z) \psi(\tilde{z}) + h_{C}(z;\lambda) \frac{\partial \tilde{z}}{\partial \lambda} \psi^{\dagger}(z) \tilde{\partial} \psi(\tilde{z}) \Big]
\end{equation}
To write the answer explicitly, we set up a few shorthands:
\begin{align}
    \Delta(z) &= (b_{1} + b_{2} - a_{1} - a_{2} ) z^{2} + 2(a_{1}a_{2} - b_{1}b_{2}) z + (a_1 + a_2) b_1 b_2 - (b_1 + b_2 ) a_1 a_2 \notag \\
    &= (b_{1} + b_{2} - a_{1} - a_{2} ) (z-z_{+}) (z-z_{-}) > 0 \\
    \gamma(z) &= (b_{1} + b_{2} - a_{1} - a_{2} ) z + a_{1}a_{2} - b_{1}b_{2} \label{nlgamma} \\
    \Pi_{a}(z) &= (a_1 -z) (a_2 - z) \qquad {\rm and} \qquad
    \Pi_{b}(z) = (b_1 - z)(b_2-z)
\end{align}
Quantity~$\Delta(z)$ vanishes when $dX(z)/dz = 0$, which happens at $z_\pm$; see equation~(\ref{defzpm}). Using this notation, as well as the $\xi$ defined in~(\ref{defxi}), varying a single endpoint gives:
\begin{align}
& \partial_{a_{i}} \hmod = \label{dahmodz} \\
& 2\pi i \int dz \frac{\Pi^{2}_{b}(z)}{\Delta^{2}(z)} 
\left( - \frac{\Pi^{2}_{a}(z)}{(a_{i}-z)^{2}} T(z) 
- \frac{ \xi \Pi_{a}(z)}{\Pi_{b}(a_{i}) \gamma(z)} \tilde{T}(z) 
+ \frac{\xi ( \Delta(z) - 2 \Pi_{b}(a_{i}) \tfrac{\Pi_{a}(z)}{(a_{i}-z)} )  }{2 \Pi_{b}(a_{i}) \gamma^{2}(z)}  
\psi^{\dagger}(z) \psi(\tilde{z})  \right) 
\nonumber \\
& \partial_{b_{i}} \hmod = \label{dbhmodz} \\
& 2\pi i \int dz \frac{\Pi^{2}_{a}(z)}{\Delta^{2}(z)} 
\left(  \frac{\Pi^{2}_{b}(z)}{(b_{i}-z)^{2}} T(z) 
- \frac{ \xi \Pi_{b}(z)}{\Pi_{a}(b_{i}) \gamma(z)} \tilde{T}(z) 
+ \frac{\xi ( \Delta(z) + 2 \Pi_{a}(b_{i}) \tfrac{\Pi_{b}(z)}{(b_{i}-z)} )  }{2 \Pi_{a}(b_{i}) \gamma^{2}(z)}  
\psi^{\dagger}(z) \psi(\tilde{z})  \right) 
\nonumber
\end{align}
Substituting these expressions in $[\ldots, \hmod]$ shows that they are modular scrambling modes:
\begin{equation}
\!\!\! \left[- \frac{1}{2 \pi i} \partial_{a_{i}} \hmod , \hmod \right] = \partial_{a_{i}} \hmod
    \qquad {\rm and} \qquad
    \left[+ \frac{1}{2 \pi i} \partial_{b_{i}} \hmod , \hmod \right] = \partial_{b_{i}} \hmod
\end{equation}
By the same token, expressions~(\ref{dahmodz}-\ref{dbhmodz}) also serve as generators of modular parallel transport, up to factors of $\pm (2 \pi i)^{-1}$. In summary, modular parallel transport in a general direction in kinematic space is given by:
\begin{equation}
V^{\text{(2-int)}}_{\lambda} = \frac{1}{2\pi i} \left[ - \frac{\partial a_{1}}{\partial \lambda} \partial_{a_{1}} \hmod + \frac{\partial b_{1}}{\partial \lambda} \partial_{b_{1}} \hmod - \frac{\partial a_{2}}{\partial \lambda} \partial_{a_{2}} \hmod + \frac{\partial b_{2}}{\partial \lambda} \partial_{b_{2}} \hmod \right]
\label{dlambdaz}
\end{equation}
These calculations confirm the conclusion, which we reached in $X$-space in equation~(\ref{dlambdax}). 

A straightforward calculation verifies that expression~(\ref{dlambdaz}) equals $(\partial_\lambda \beta) \beta^{-1} \hmod$. This confirms the guess of equation~(\ref{mptderivative}) and the other points highlighted in Section~\ref{sec:1int}. 

\subsection{Relation to modular inclusion}
Formula~(\ref{dlambdaz}) is simple whereas explicit expressions like (\ref{dahmodz}-\ref{dbhmodz}) are quite involved. There must be a reason for the simplicity of the final answer, when it is expressed in terms of partial derivatives of $\hmod$. That reason is that the algebras of observables in regions $(a_1, b_1-db_1) \cup (a_2, b_2)$ and $(a_1, b_1) \cup (a_2, b_2)$ form a half-sided modular inclusion \cite{Wiesbrock:1992mg,Borchers:1999}.

Consider two algebras of observables $\mathcal{B} \subset \mathcal{A}$, which act on the same Hilbert space $\mathcal{H}$. Take a state $|0\rangle \in \mathcal{H}$, which is (i) cyclic and (ii) separating in both $\mathcal{A}$ and $\mathcal{B}$. These conditions mean that: (i) states of the form $a |0\rangle$, taken over $a \in \mathcal{B}$ (the smaller of the two algebras), span the whole Hilbert space; and (ii) that $a |0\rangle$ is only zero if $a = 0$. Consider the modular flow
\begin{equation} 
\sigma^{t}_{\mathcal{A}}(\cdot) = e^{it \hmod_{\mathcal{A}}} (\cdot )\, e^{-it \hmod_{\mathcal{A}}},
\label{modflow}
\end{equation} 
where $\hmod_{\mathcal{A}}$ is the modular Hamiltonian in state $|0\rangle$ computed in the larger algebra $\mathcal{A}$. If the subalgebra $\mathcal{B}$ is closed under this flow for all positive times, that is if
\begin{equation}
	\sigma^{t}_{\mathcal{A}}(\mathcal{B}) \subset \mathcal{B} \qquad \forall~t\geq 0,
	\label{modincl}
\end{equation}
then $\mathcal{B} \subset \mathcal{A}$ are said to form a half-sided modular inclusion \cite{Wiesbrock:1992mg,Borchers:1999}. Whenever it occurs, the half-sided modular inclusion is a delicate circumstance: (\ref{modincl}) concerns evolving $\mathcal{B}$ with the modular operator of the larger algebra $\mathcal{A}$, which generically spills out of $\mathcal{B}$ into $\mathcal{A}$. For our purposes, it is crucial that modular inclusion implies the following commutator relation between the modular Hamiltonians \cite{Casini:2017roe}:
\begin{equation}
[\hmod_{\mathcal{A}} - \hmod_{\mathcal{B}}, \hmod_{\mathcal{B}}] = 2\pi i (\hmod_{\mathcal{A}} - \hmod_{\mathcal{B}})
\label{inclcomm}
\end{equation}
Applied to planar and spherical regions in vacuum states of holographic conformal field theories, this relation explains the emergence of Poincar{\'e} symmetry in the bulk AdS geometry \cite{Casini:2017roe,mpt,scramblingmodes,Leutheusser:2021frk}.
By essentially the same reasoning, it also explains formula~(\ref{dlambdaz}). 

Let $\mathcal{A}$ be the algebra of observables on the double interval region $(a_1, b_1) \cup (a_2, b_2)$ and $\mathcal{B}$ that on \mbox{$(a_1, b_1-db_1) \cup (a_2, b_2)$}. In that case, $\hmod_{\mathcal{A}} - \hmod_{\mathcal{B}} = \partial_{b_1} \hmod db_1$. We are working in the left-moving coordinate $z = x - t$ so this setup satisfies $\mathcal{B} \subset \mathcal{A}$. If modular inclusion can be established for these algebras then equation~(\ref{inclcomm}) will become
\begin{equation}
\left[ \frac{1}{2\pi i}\, \partial_{b_1} \hmod, \hmod\right] = \partial_{b_1} \hmod
\end{equation}
and $(2 \pi i)^{-1} \partial_{b_1} \hmod$ is the solution of the modular parallel transport problem (\ref{mptdef1}-\ref{mptdef2}). In particular, as an eigen-operator of $[\ldots, \hmod]$, $(2 \pi i)^{-1} \partial_{b_1} \hmod$ is already zero mode-free. 

\paragraph{Verifying modular inclusion} 

\begin{figure}[tbp]
\centering 
\includegraphics[width=.8\textwidth]{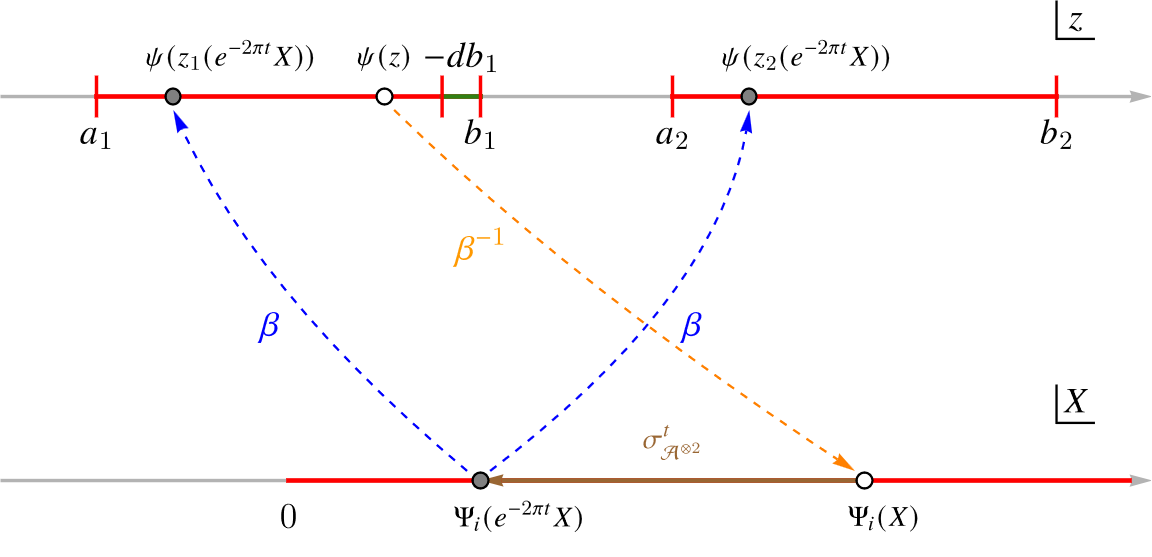}
\caption{Equation~(\ref{modflowdetail}), represented graphically.}
\label{fig:2-inclusion}
\end{figure}

Consider a fermion $\psi(z)$ from the smaller region, that is $(a_1, b_1-db_1) \cup (a_2,b_2)$. To establish modular inclusion, it suffices to show that the fermion remains there under modular flow~(\ref{modflow}) of the bigger region, for positive $t$. Here we use the fact that the isomorphism $\beta$ intertwines modular flows in $z$-space and $X$-space \cite{Tedesco:2014eaz}. By equation~(\ref{ittflow}), $\sigma^t_{\mathcal{A}}$ sends $\psi(z)$ to:
\begin{align}
\psi(z)
\,\xrightarrow{\beta_{(a_1, b_1) \cup (a_2, b_2)}^{-1}}\, &
\,\#_1 \Psi_1(X(z)) + \#_2 \Psi_2(X(z)) 
\nonumber \\ 
\,\xrightarrow{~\sigma^t_{\!\!\mathcal{A}^{\otimes 2}(I)~}}\, & 
\,\#'_1 \Psi_1(e^{-2\pi t}X(z)) + \#'_2 \Psi_2(e^{-2\pi t} X(z)) 
\label{modflowdetail} \\
\,\xrightarrow{\beta_{(a_1, b_1) \cup (a_2, b_2)}\phantom{|}}\, &
\,\#''_1\, \psi(z_1(e^{-2\pi t}X(z))) + \#''_2\, \psi(z_2(e^{-2\pi t}X(z))) 
\nonumber
\end{align}
The steps are represented graphically in Figure~\ref{fig:2-inclusion}.
 Here $X$ is defined for the bigger region $(a_1, b_1) \cup (a_2, b_2)$---precisely as in equation~(\ref{xzmapping}). The successive steps use equations~(\ref{defbetainverse}), (\ref{defboost}), and (\ref{finalbeta}). We suppressed detailed coefficients, writing them schematically as $\#$s. 

The two inverses $z_{1,2}(X)$---see equation~(\ref{defz12})---map $X \in (0, \infty)$ to $z_1(X) \in (a_1, b_1)$ and $z_2(X) \in (a_2, b_2)$, respectively. Therefore, the only thing to check is that
\begin{equation}
z \in (a_1, b_1 - db_1)~\Longrightarrow~z_1(e^{-2\pi t} X(z)) \in (a_1, b_1-db_1) \qquad \forall t \geq 0
\end{equation}
This follows because $z_1(X(z)) = z$ in this range and $z_1(X)$ is a monotonic function of $X$. 

\paragraph{Other terms} The above analysis referred to variations of double interval regions in one specific lightcone direction: $b_1$, i.e. the left-moving coordinate of the right endpoint of the first interval. For $b_2$---the left-moving coordinate of the right endpoint of the second interval---the argument is identical. 

We now turn to variations in left endpoints, taking $a_2$ for a start. Let $\mathcal{A}$ be the algebra of observables on the complement of $(a_1, b_1) \cup (a_2, b_2)$, and $\mathcal{B}$ be the algebra on the complement of $(a_1, b_1) \cup (a_2 - da_2, b_2)$. By equation~(\ref{hmodschem}), working with complements sends $\hmod \to -\hmod$. If modular inclusion holds for $\mathcal{B} \subset \mathcal{A}$, we will therefore have:
\begin{equation}
\left[ -\frac{1}{2\pi i}\, \partial_{a_2} \hmod, \hmod\right] = \partial_{a_2} \hmod
\label{modincla}
\end{equation}
As before, this means that $-(2 \pi i)^{-1} \partial_{a_2} \hmod$ solves the modular parallel transport problem for variations in $a_2$. 

To verify modular inclusion for the complementary regions, we again inspect the modular flow. It follows structure~(\ref{modflowdetail}) except now $X$ gets dressed by a factor $e^{2\pi t}$ instead of $e^{-2\pi t}$. Assume that the endpoints are properly ordered ($a_2 < b_1$) and that the inverses $z_{1,2}$ are correctly indexed. (This means that $z_2(X(z)) = z$ for $z \in (b_1, a_2)$; the alternative would be $z = z_1(X(z))$.) For the half-sided modular inclusion, we now need:
\begin{equation}
z \in (b_1, a_2 - da_2)~\Longrightarrow~z_2(e^{2\pi t} X(z)) \in (b_1, a_2-da_2) \qquad \forall t \geq 0
\end{equation} 
In this range $X(z) < 0$ so the conclusion again follows from $z_2(X(z)) = z$ and from monotonicity of $z_2(X)$ in $X$.

The argument for variations in $a_1$ is the same except that it uses the other inverse $z_1(X)$. That inverse sends $X \in (-\infty, 0)$ to $z \in (-\infty, a_1) \cup (b_2, +\infty)$ so it is not strictly speaking monotonous. This introduces some tediousness in the last step but does not affect the final conclusion. 

Finally, we briefly comment on the right-moving sector. There, the signs $\pm (2 \pi i)$ are reversed because boosts transform $X \to e^{-2\pi t} X$ but $\bar{X} \to e^{2\pi t} \bar{X}$. As a result, we find a half-sided modular inclusion for variations in the right-moving coordinates of left endpoints: $\bar{a}_1$ and $\bar{a}_2$. For variations in $\bar{b}_1$ and $\bar{b}_2$, we have a modular inclusion for complementary regions. The overall pattern in the right-movers is opposite to the left-movers, as is mandated by time reversal symmetry. 

\section{Interesting limits}
\label{sec:limits}
In this section we consider a few special limits of our results. We present illustrative plots of the smearing functions for the local and non-local stress tensor term as well as the non-local current. 

Before doing so, we need a few preliminary comments. Deforming the two-interval region $(a_1, b_1) \cup (a_2, b_2)$ is equivalent to deforming its complement $(-\infty, a_1) \cup (b_1, a_2) \cup (b_2, \infty)$. Because we are working in the vacuum of a conformal field theory, the latter is also conformally equivalent to a two-interval region. This appears to suggest that our results should be symmetric under the exchange of left endpoints $a_1$ and $a_2$ with right endpoints $b_1$ and $b_2$. As we shall see, this tentative conclusion is basically correct but subtle.

This is best seen with reference to formula~(\ref{dahmodz}). The smearing functions of the non-local terms feature in their denominators the quantity $\gamma(z)$, which was defined in (\ref{nlgamma}). When $\gamma(z) = 0$, these smearing functions diverge. The divergence appears at
\begin{equation}
z_* = \frac{a_1 a_2 - b_1 b_2}{a_1 + a_2 - b_1 - b_2},
\end{equation}
which is invariant when we swap the $a$'s with the $b$'s. Assuming the endpoint values are distinct and properly ordered ($a_1 < b_1 < a_2 < b_2$), we have $z_* \in (b_1, a_2)$. This is easy to understand: the point $z=z_*$ is paired with infinity under the map (\ref{defztilde}): $\tilde{z}(z_*) = \infty$. Because conformal transformations shift infinity around, the singularity in (\ref{dlambdaz}) has no invariant meaning. Consequently, comparing different instances of modular parallel transport requires special care. In particular, we should not expect the plots of smearing functions to appear symmetric between $(a_1, b_1) \cup (a_2, b_2)$ and its complement. To account for this subtlety, each limit considered below will be implemented in two ways: with respect to $I_1 = (a_1, b_1)$ and with respect to the `middle' interval $(b_1, a_2)$, which contains $z_*$. 

In all plots, we display the generator $V_{a_1}$ and use blue, orange, green, red to represent successive stages in which a limit is approached.

\subsection{One interval shrinks}
The first limit we consider is to shrink one interval. As explained above, we have two ways to do that: (i) shrink one of $I_1 = (a_1, b_1)$ or $I_2 = (a_2, b_2)$ (say, $b_2 \to a_2$), or (ii) shrink the interval between them ($b_1 \to a_2$). For the first case, we will take $I_1 = (a_1, b_1) = (-1, 0)$ and consider a few shrinking choices of the second interval: starting with $I_2 = (a_2, b_2) = (10,11)$, then letting $b_2 = 10.1, 10.01, 10.001$. For the second case, we take a set-up symmetric about $z=0$: we fix $a_{1} = -5$ and $b_{2} = 5$, then let $-b_{1} = a_2 = 1, 0.1, 0.01, 0.001$. In order to highlight features which are not determined by conformal kinematics, we arranged for these data points to have the same conformal cross-ratios
\begin{equation}
\chi = \frac{(a_{2} - a_{1})(b_{2} - b_{1})}{(a_{2} - b_{1})(b_{2} - a_{1})}
\end{equation}
They take values $1+10^{-2}$, $1+10^{-3}$, $1+10^{-4}$ and $1+10^{-5}$ and the shrinking limit is $\chi \to 1$. 

\begin{figure}[tbp]
\centering 
\includegraphics[width=.4\textwidth]{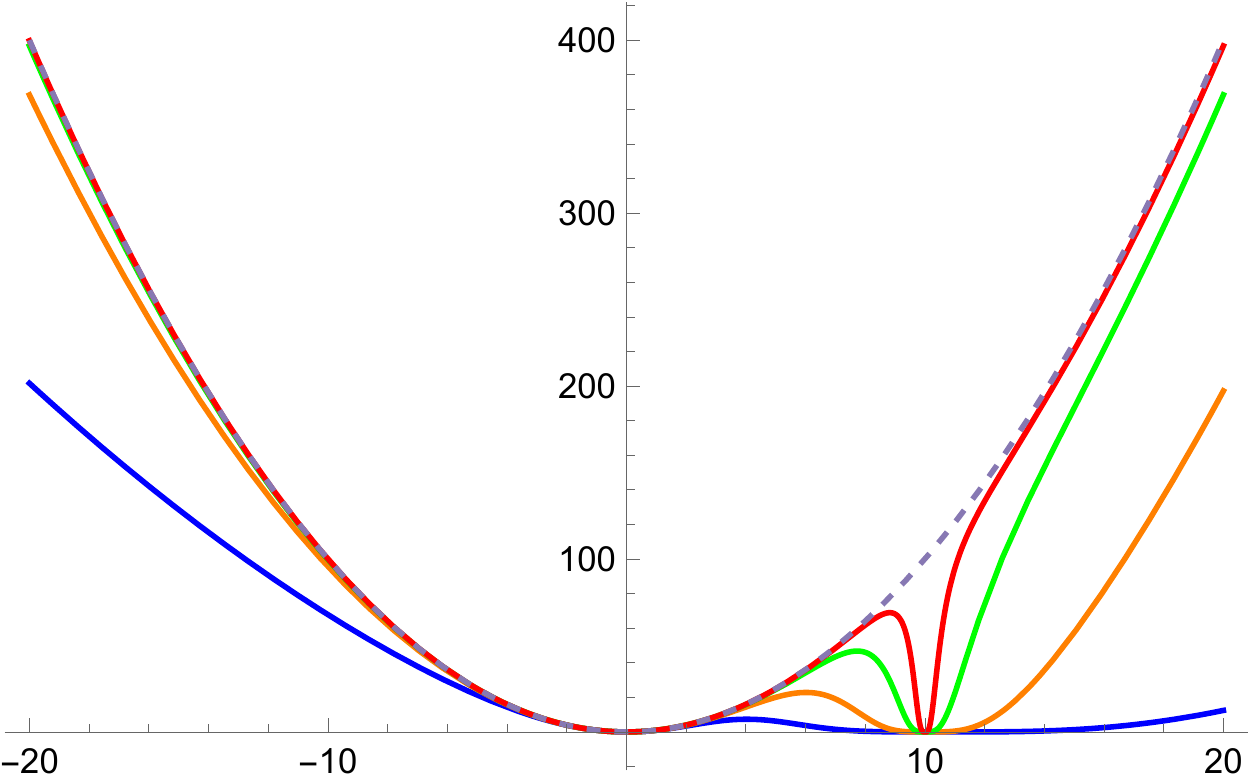}
\hfill
\includegraphics[width=.4\textwidth]{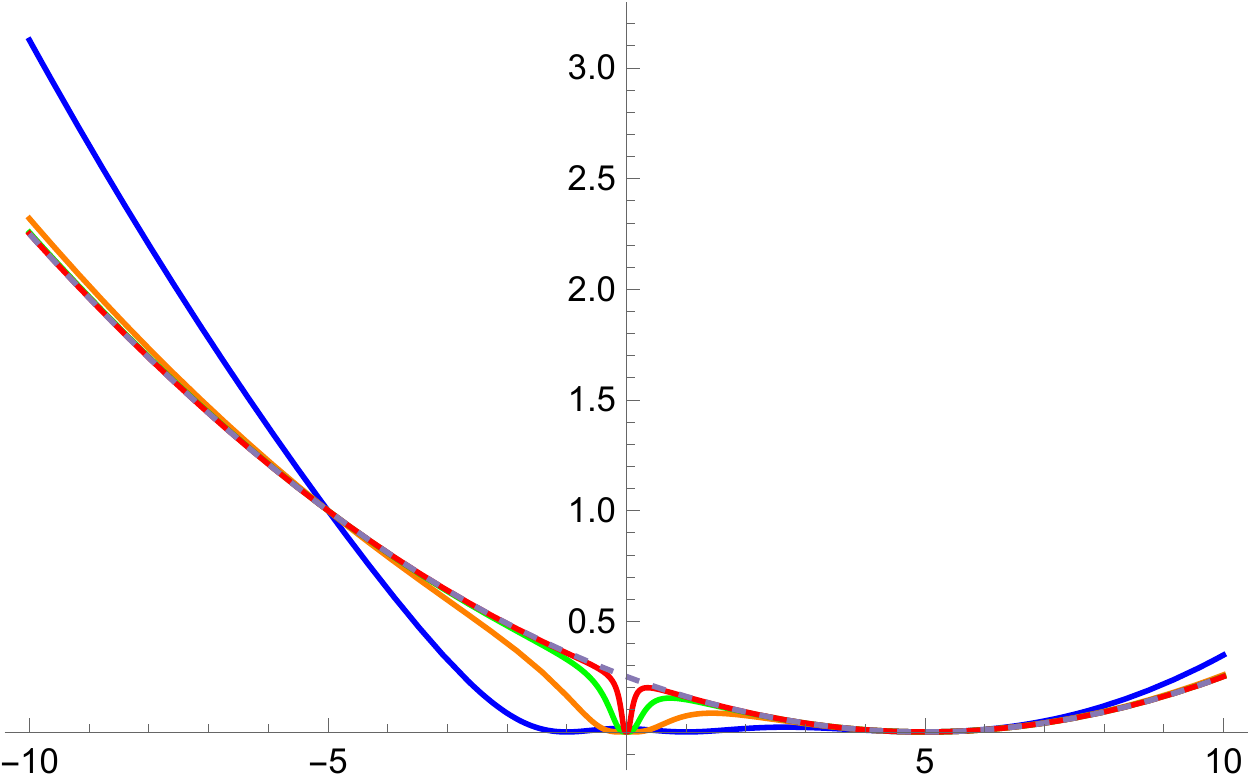}
\caption{The smearing function of the local stress tensor in the $V_{a_{1}}$ generator of modular parallel transport. On the left panel, we took $I_1 = (-1, 0)$ and $I_2 = (10, b_2)$, with $b_2 = 11, 10.1, 10.01, 10.001$. In the limit $b_2 \to 10$, the smearing function goes to the one-interval answer (dashed line) except at $z = 10$. On the right panel, we took $a_{1} = -5$ and $b_{2} = 5$ with $a_{2} = -b_{1} = 1, 0.1, 0.01, 0.001$. Similarly, the smearing function also approach the one-interval answer with a dip. The dip now locates at $z=0$, where $b_{1}$ and $a_{2}$ meet.}
\label{fig:shrink1int-local}
\end{figure}

\paragraph{Local stress tensor}
The smearing function for the local stress tensor does not have $\gamma(z)$ in its denominator. Therefore, it is divergence-free and we can safely plot it on the whole real line; see Fig~\ref{fig:shrink1int-local}. Cases (i) and (ii) behave in the same way in that the smearing function approaches the single-interval answer: $(a_1, b_1)$ in case~(i) and $(a_1, b_2)$ in case~(ii). The convergence holds everywhere except the point $z = a_2$ to which one interval shrinks, where the smearing function vanishes. The convergence is therefore point-wise, not uniform, which means that modular parallel transport `remembers' about the second interval even when the latter shrinks to a point.

Algebraically, we can see this by noting that the $T(z)$ smearing function for moving $a_1$ must vanish at the remaining three interval endpoints: $b_1$, $a_2$, and $b_2$. (In the one-interval case, the smearing function in $V_{a}^{\rm (1-int)}$ vanishes at $b$ and the smearing function in $V_{b}^{\rm (1-int)}$ vanishes at $a$.) The vanishing at $z=a_2$ is a feature, which distinguishes the two-interval answer~(\ref{dlambdaz}) from the single-interval answer~(\ref{1intfermion}) even when the second interval vanishes. More geometrically, shifting $a$ in the single interval case is generated by a dilation about $z=b$. But in the two-interval case, the $T(z)$ smearing function for moving $a_1$ is not quadratic, so it does not generate a global conformal transformation.

\begin{figure}[tbp]
\centering 
\includegraphics[width=.4\textwidth]{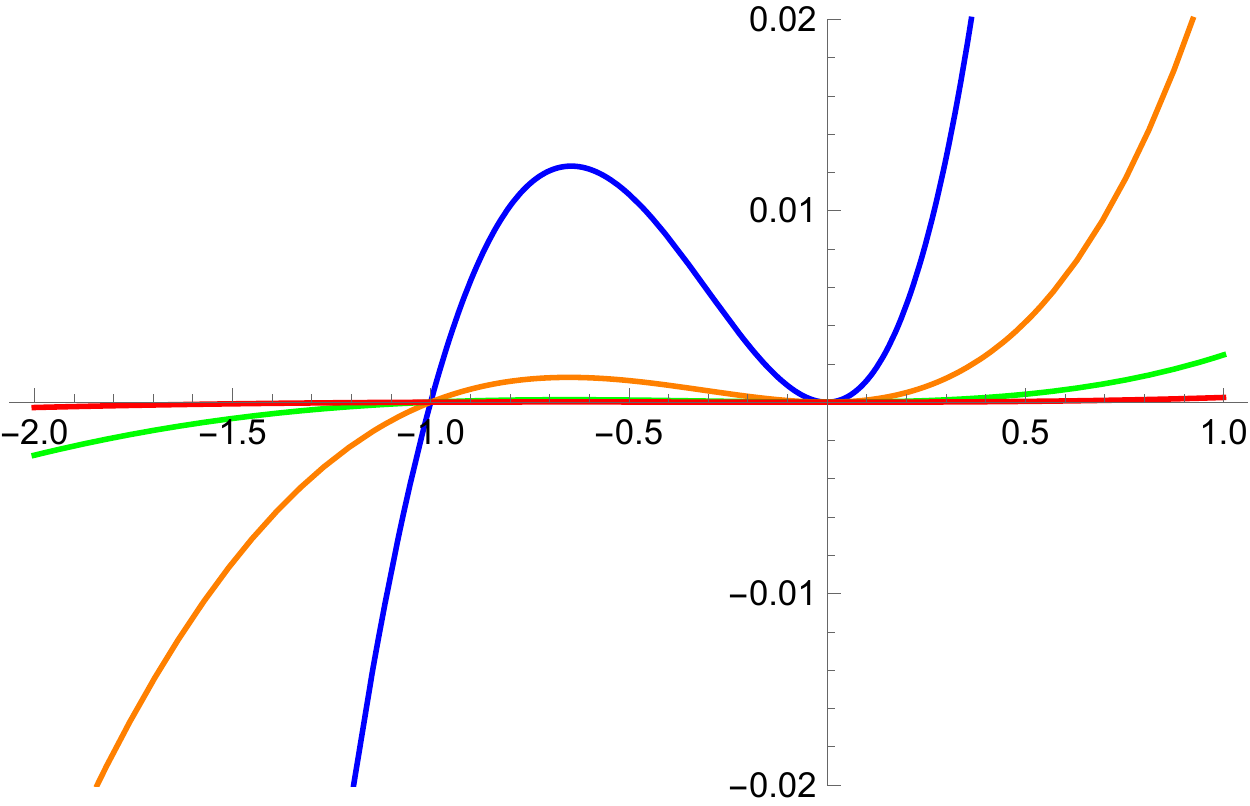}
\hfill
\includegraphics[width=.4\textwidth]{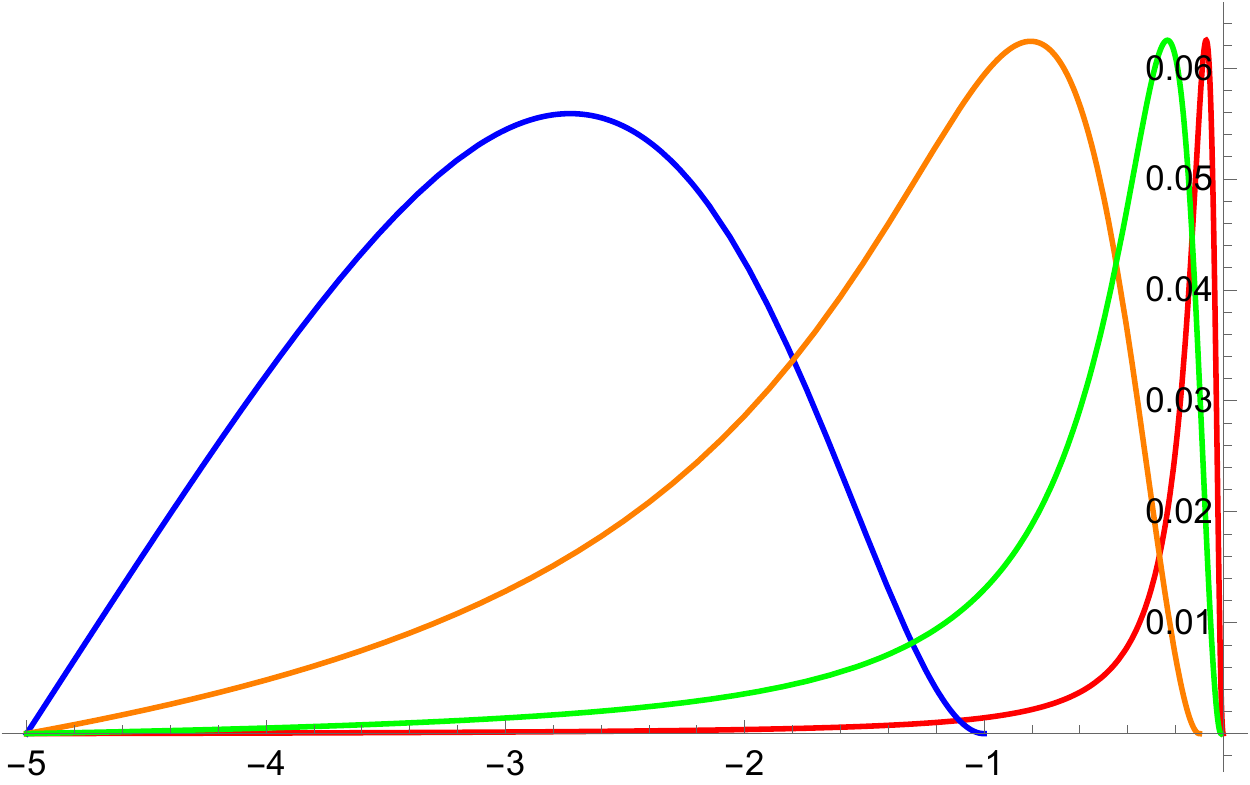}
\caption{The smearing function of the non-local stress tensor in the $V_{a_{1}}$ generator of modular parallel transport. In the left panel, we took $I_1 = (-1, 0)$ and $I_2 = (10, b_2)$, with $b_2 = 11, 10.1, 10.01, 10.001$. As we shrink the interval, the smearing function inside $I_{1}$ vanishes. In the right panel, we took $a_{1} = -5$ and $b_{2} = 5$ with $a_{2} = -b_{1} = 1, 0.1, 0.01, 0.001$. In this case, a peak is always present inside $I_1$. A similar peak also appears in $I_{2}$ (not shown).}
\label{fig:shrink1int-nonlocal}
\end{figure}

\paragraph{Non-local stress tensor}
The smearing functions for cases (i) and (ii) are shown in Figure~\ref{fig:shrink1int-nonlocal}. They share several common features:
\begin{itemize}
\item They vanish at all interval endpoints, including the endpoint $a_1$, which is being shifted by $V_{a_1}$. The zeroes at left endpoints $z = a_1, a_2$ are first order whereas the zeroes at right endpoints $z = b_1, b_2$ are second order.
\item The smearing function has local maxima inside $I_1 = (a_1, b_1)$ and $I_2 = (a_2, b_2)$. 
\item The smearing function diverges at $z_* \in (b_1, a_2)$.
\item Away from the divergence, the smearing function approaches zero point-wise in the shrinking limit.
\end{itemize}
One thing that distinguishes cases (i) and (ii) is the behavior of the peaks in intervals $I_1$ and $I_2$. The peak magnitudes of the smearing function tend to zero in case (i) (shrinking interval $I_2$), but when the middle interval $(b_1, a_2)$ shrinks (case (ii)) the peaks remain finite. The point-wise convergence to zero obtains because the peak locations approach the shrunk interval at $z = a_2$. 

Same as in the smearing function for the local stress tensor, the presence of the singularity distinguishes the transport of the double-interval from the single interval case even if one of the constituent intervals shrinks to zero size. The fermions of the two intervals always remain coupled.

\begin{figure}[t]
\centering 
\includegraphics[width=.4\textwidth]{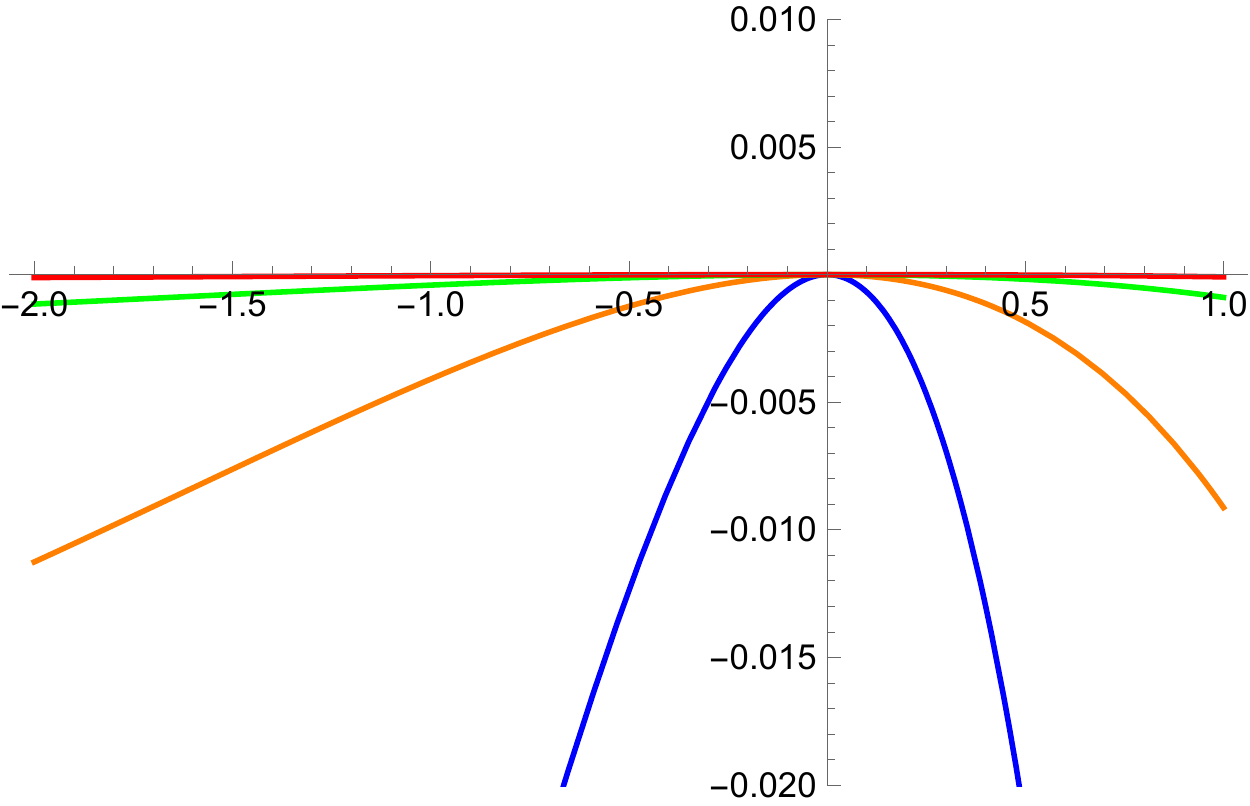}
\hfill
\includegraphics[width=.4\textwidth]{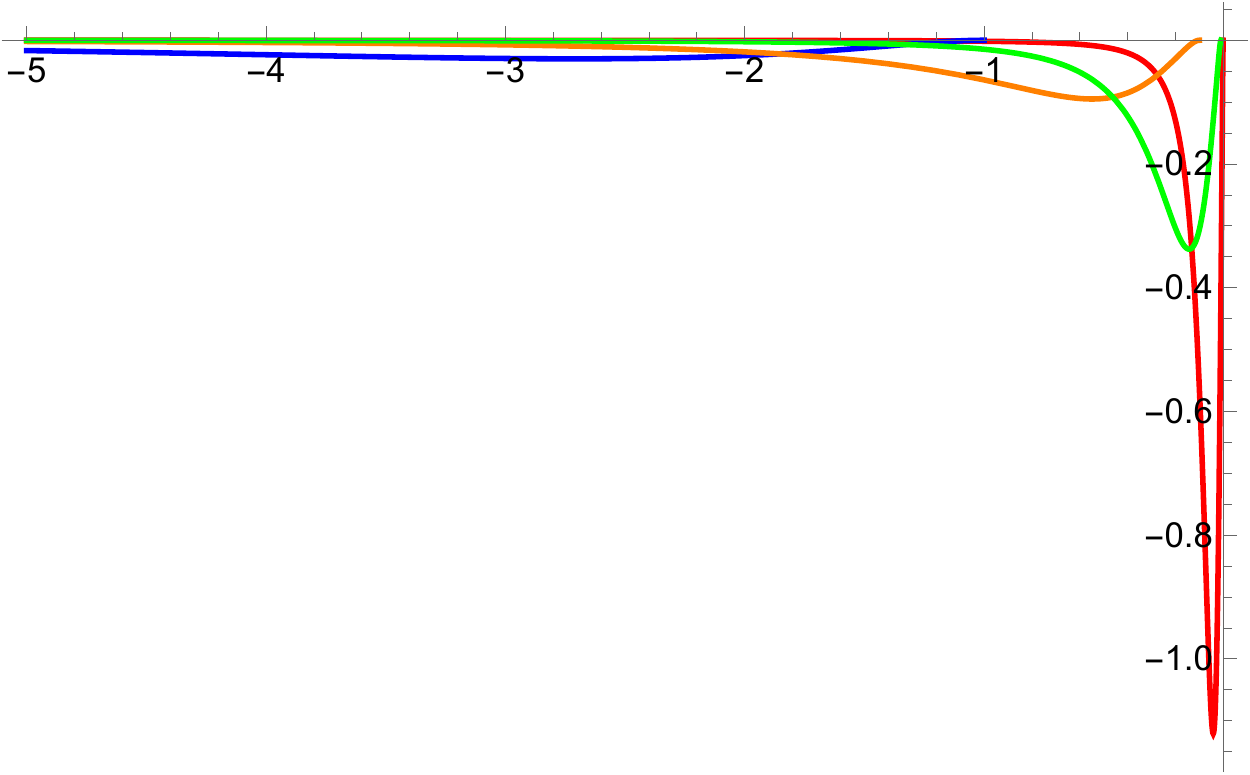}
\caption{The smearing function of the non-local fermion current in the $V_{a_{1}}$ generator of modular parallel transport. In the left panel, we took $I_1 = (-1, 0)$ and $I_2 = (10, b_2)$, with $b_2 = 11, 10.1, 10.01, 10.001$. In the right panel, we took $a_{1} = -5$ and $b_{2} = 5$ with $a_{2} = -b_{1} = 1, 0.1, 0.01, 0.001$.}
\label{fig:shrink1int-current}
\end{figure}

\paragraph{Non-local fermion current}
The smearing functions are shown in Figure~\ref{fig:shrink1int-current}. Once again, in the shrinking interval limit, they approach zero point-wise everywhere except the singularity at $z = z_*$. 

\subsection{Two intervals are taken farther apart}
\begin{figure}[tbp]
\centering 
\includegraphics[width=.4\textwidth]{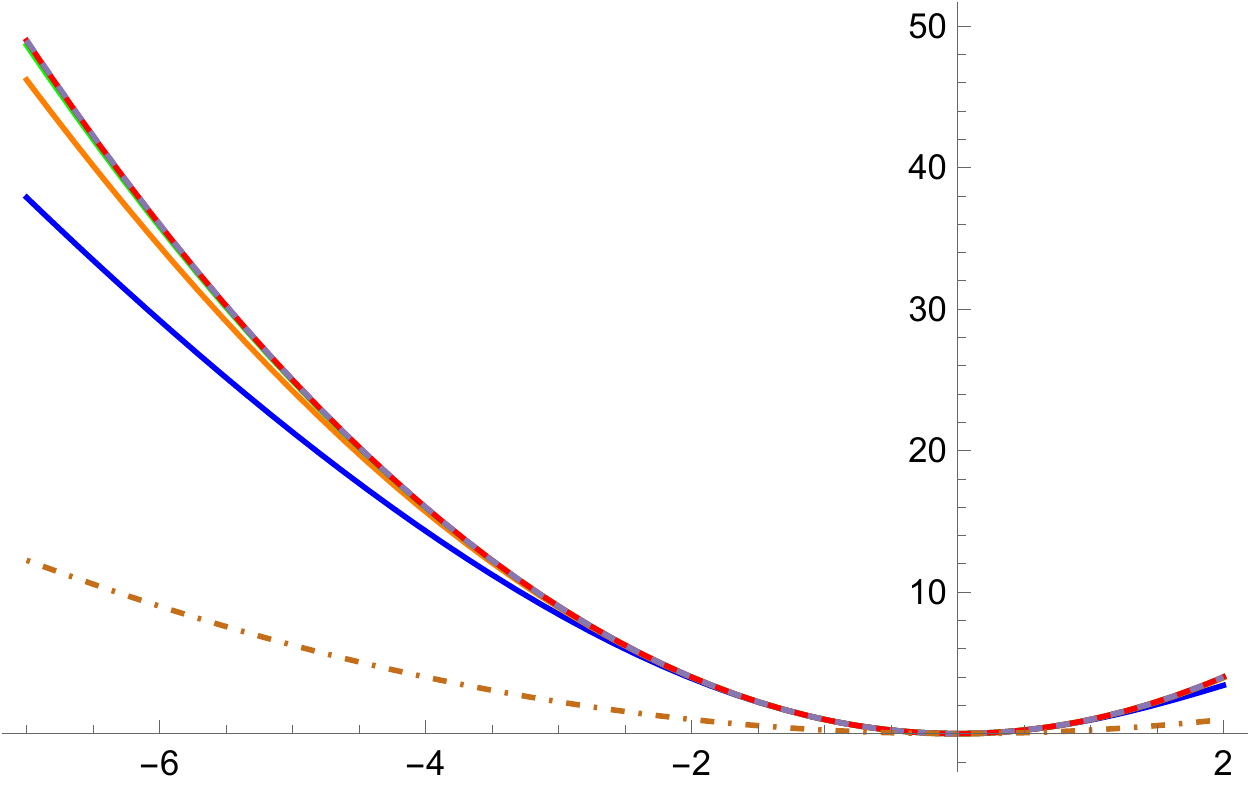}
\hfill
\includegraphics[width=.4\textwidth]{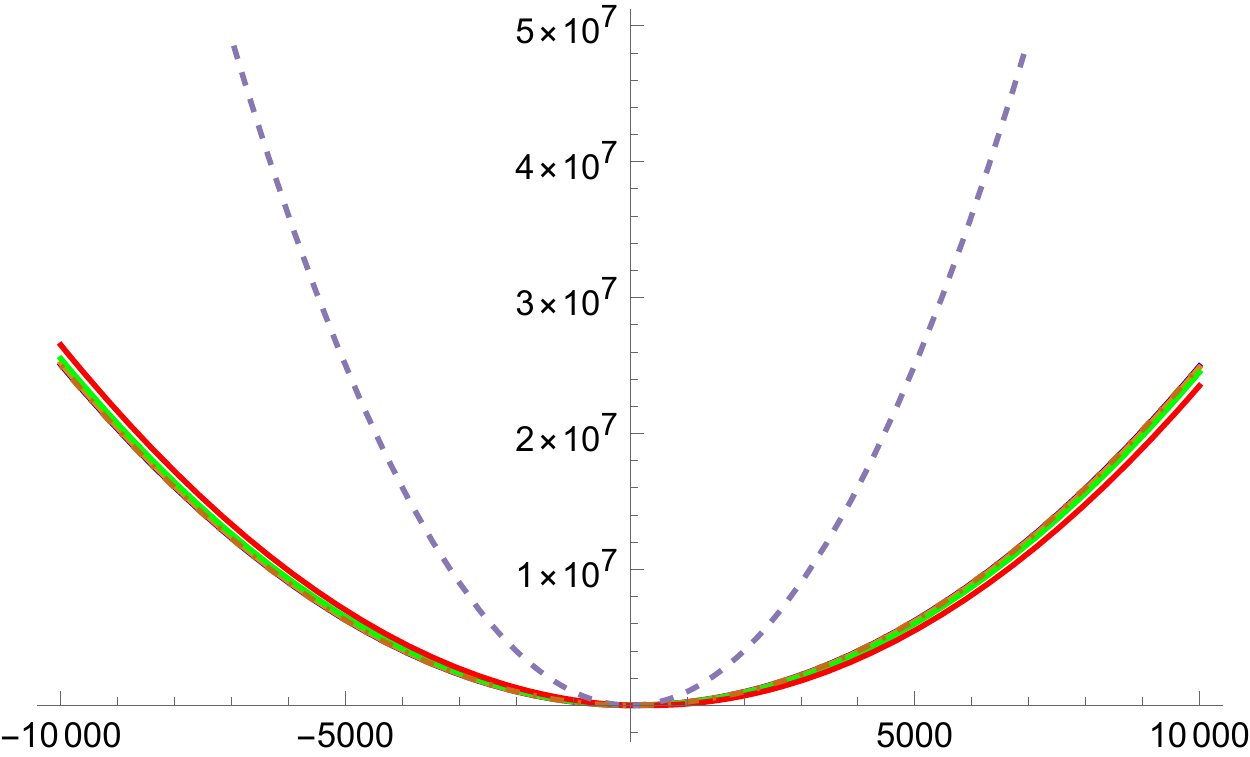}
\caption{The smearing function of the local stress tensor in the $V_{a_{1}}$ generator of modular parallel transport. We fix the sizes of both intervals and increase the distance between them. In the left panel, we focus on the behavior around interval $I_1$. In the right panel, we extend the plot to a much larger range. The dashed line is the na{\"\i}ve one-interval answer for $(a_{1},b_{1}) = (-1,0)$; it is the large separation limit in the left panel. The dot-dashed line is the one-interval solution for $(a_1, b_2)$; it is the large separation limit in the right panel.}
\label{fig:shrink2int}
\end{figure}

We fix the sizes of $I_1$ and $I_2$ and increase the distance between them. We take $I_1 = (a_1, b_1) = (-1, 0)$ and consider a series of unit-size intervals $I_2 = (a_2, b_2) = (a_2, a_2 + 1)$ with $a_2 = 10, 30, 100, 300$. The $a_2$s are chosen to yield the same conformal cross-ratios as before: $\chi = 1+10^{-2}$, $1+10^{-3}$, $1+10^{-4}$ and $1+10^{-5}$. We again inspect $V_{a_1}$ and find the following:

\begin{itemize}
	\item The smearing function for the local stress tensor is shown in two plots in Fig~\ref{fig:shrink2int}. The first one focuses on the behaviour around interval $I_{1}$. It is clear that as the distance between $I_{1}$ and $I_{2}$ gets larger, the smearing function approaches the na{\"\i}ve one-interval solution for interval $I_{1}$ alone. On the other hand, at scales much larger than the separation between $I_1$ and $I_2$, the smearing function approaches the one-interval form for $(a_1, b_2)$, that is the union of $I_1$ and $I_2$ and everything in between. Thus, in its local stress tensor component, the modular parallel transport of double intervals interpolates between two distinct single-interval answers. 
	\item Non-local contributions. Both of them---the non-local stress tensor and the current---exhibit similar behaviors. Their smearing functions approach zero on the small intervals $I_{1,2}$ but grow in $(b_1, a_2)$ and the `interval' that contains infinity.
\end{itemize}

\paragraph{A counterpart: shrinking two neighboring intervals}
We promised to take every limit in two ways, applying analogous deformations to $I_{2}$ and to the interval between $I_1$ and $I_2$. Setting $I_{1,2}$ far apart is related by a conformal transformation to shrinking them simultaneously. It now behooves us inspect a limit in which, alongside $I_1$, the middle interval $(b_1, a_2)$ is shrunk instead of $I_2$. We will be sending $b_{1,2} \rightarrow a_{2}$. 

In this limit, we recover the modular parallel transport of a single interval $(a_1, a_2)$. 

\subsection{Comparison with holography}
One motivation for studying these special examples in the free fermion is a comparison with holographic theories. There, in the limit of large central charge, modular parallel transport can be holographically interpreted as a bulk diffeomorphism, which maps the corresponding Ryu-Takayanagi surfaces to one another \cite{mpt}. While modular parallel transport in the free fermion theory has no such interpretation, it is useful as a computable, small central charge counterpoint to holographic examples.  

In holography, two-interval regions can be categorized into two classes: the connected phase and the disconnected phase. This refers to whether their entanglement wedges in the bulk are connected or disconnected. The transition, which happens when $(a_2 - b_2) (a_1 - b_1) = (a_2 - b_1) (a_1 - b_2)$, is sharp at infinite $N$ but smoothed out by enhanced $1/N$ corrections \cite{dongwang}. Away from the transition point, modular parallel transport in holographic theories decouples, from the bulk point of view, into two separate single-interval problems. This is because the relevant Ryu-Takayanagi surfaces are pairs of distant geodesics, each of which is independently deformed under bulk modular transport. 

Of course, the modular parallel transport of double intervals in the free fermion theory does not decompose in this way. On the other hand, the modular parallel transport of single intervals is the same in all 1+1-dimensional conformal field theories, holographic or otherwise, because it is fixed by conformal symmetry. For this reason, we have been interested in limits in which modular parallel transport of double intervals might reduce to that of single intervals. The extent to which this does not happen is a signature of the non-holographic character of the free fermion. 

In the end, we found the following:
\begin{itemize}
\item The only limit in which the modular parallel transport of double intervals reduces to that of a single interval is when two neighboring intervals---say, $I_2$ and the middle interval $(b_1, a_2)$---are simultaneously shrunk to zero size. This is a trivial case, in the sense that it literally reduces the problem to one interval and a complement.  
\item In all other cases, modular parallel transport of $I_1 \cup I_2$ is qualitatively distinct from the modular parallel transport of $I_1$ alone. For example, when one interval is shrunk non-local terms continue to couple the vestigial point-like interval to its partner.
\item If we focus on the local stress tensor term alone, we see that it does approach the single-interval form, albeit with subtleties. When intervals $I_{1,2}$ are taken far apart, it approaches both single-interval answers (for $I_1$ and for $I_1 \cup (b_1, a_2) \cup I_2$), but at different scales.
\end{itemize}

\section{Curvature} 
\label{sec:curvature}
This section computes the curvature of the connection, which defines modular parallel transport. As usual, the curvature two-form is given by the commutator of two parallel transport generators understood as operator-valued one-forms:
\begin{equation}
K_{\lambda \sigma} d\lambda d\sigma = \left[ V_{\lambda} d\lambda, V_{\sigma} d\sigma \right] 
\label{eq:curvature}
\end{equation}
In equation~(\ref{dlambdaz}), we found that in the free fermion theory modular transport in the direction of shifting a single endpoint is given by $\pm \partial \hmod$. Therefore, if we take $\lambda$ and $\sigma$ in equation~(\ref{eq:curvature}) to denote $z$-space locations of interval endpoints, we will find: 
\begin{equation}
K_{\lambda \sigma} = P^0 [\pm \partial^2_{\lambda \sigma} \hmod]
\qquad\textrm{(where $\lambda, \sigma$ denote interval endpoint locations)}
\label{curv-form}
\end{equation}
Here $P^0$ is the projector onto zero-modes of $\hmod$. To explain the projector, we remind the reader that the gauge symmetry which underlies modular parallel transport comprises zero modes of $\hmod$. The curvature of a gauge bundle is of course valued in its gauge group.  

The commutator in equation~(\ref{curv-form}) is straightfoward to compute directly in $z$-space, but the result takes a rather messy form, which is difficult to interpret. The calculations are simpler to conduct and present in $X$-space. A priori, six different commutators need to be computed. However, the zero mode projector in (\ref{curv-form}) sets $K_{a_1 a_2} = K_{b_1 b_2} = 0$. This is because $V_{a_1}$ and $V_{a_2}$ (respectively $V_{b_1}, V_{b_2}$) are both scrambling modes of modular frequency $-2\pi i$ (resp. $+2\pi i$), so their commutator is a modular mode of frequency $-4 \pi i$ (resp. $+4\pi i$). The projector in (\ref{curv-form}) annihilates it. 

We are left with four non-vanishing components of the curvature two-form. We use the results of Section~\ref{sec:commutators} and the notation from (\ref{zerocurrent}-\ref{zerostress}), complemented by a new zero mode that we have not encountered previously:
\begin{align}
    Q^{(2)}_{21} &= \int dX X^{2} \left( \partial \Psi^{\dagger}_{2} \partial \Psi_{1} - \partial \Psi^{\dagger}_{1} \partial \Psi_{2} \right) 
\end{align}
The non-vanishing curvature two-form components are:
\begin{align}
\!\!\!\! \left[ V_{a_{1}} , V_{b_{1}}  \right] &= -\frac{b_{2}-a_{2}}{(b_{1} - a_{1}) \sqrt{\xi} } \left( Q^{(2)}_{21} - Q^{(1)}_{12} - Q^{(1)}_{21} + \frac{1}{2} \left( Q^{(0)}_{12} - Q^{(0)}_{21}   \right)  \right) - \frac{2}{(b_{1} - a_{1})^{2}} Q^{(1)}_{22} 
\label{curv1} \\
\!\!\!\! \left[ V_{a_{1}} , V_{b_{2}}  \right] &= -\frac{b_{1}-a_{2}}{(b_{2} - a_{1}) \sqrt{\xi} } \left( Q^{(2)}_{21} + Q^{(1)}_{12} + Q^{(1)}_{21} + \frac{1}{2} \left( Q^{(0)}_{12} - Q^{(0)}_{21}   \right)  \right) - \frac{2}{(b_{2} - a_{1})^{2}} Q^{(1)}_{11} \\
\!\!\!\!  \left[ V_{a_{2}} , V_{b_{1}}  \right] &= -\frac{b_{2}-a_{1}}{(b_{1} - a_{2}) \sqrt{\xi} } \left( Q^{(2)}_{21} - Q^{(1)}_{12} - Q^{(1)}_{21} + \frac{1}{2} \left( Q^{(0)}_{12} - Q^{(0)}_{21}   \right)  \right) - \frac{2}{(b_{1} - a_{2})^{2}} Q^{(1)}_{22} \\
\!\!\!\!  \left[ V_{a_{2}} , V_{b_{2}}  \right] &= -\frac{b_{1}-a_{1}}{(b_{2} - a_{2}) \sqrt{\xi} } \left( Q^{(2)}_{21} + Q^{(1)}_{12} + Q^{(1)}_{21} + \frac{1}{2} \left( Q^{(0)}_{12} - Q^{(0)}_{21}   \right)  \right) - \frac{2}{(b_{2} - a_{2})^{2}} Q^{(1)}_{11}
\label{curv4}
\end{align}
We remind the reader that quantity $\xi$ was defined in equation~(\ref{defxi}). 

The appearance of $Q^{(1)}_{22}$ in terms involving $b_1$ and of $Q^{(1)}_{11}$ in terms involving $b_2$ may seem counterintuitive. This apparent asymmetry between left and right endpoints is an artefact of working in the $X$ coordinate, which itself treats the $a_i$'s and $b_j$'s differently. We have verified that the relevant expressions in $z$-space favor neither left nor right endpoints. 

\paragraph{Comparison with other setups}
Note how answers (\ref{curv1}-\ref{curv4}) differ from the curvature of the modular parallel transport of a single interval, which is shared with holographic theories. There, the only zero modes present in the curvature are the modular Hamiltonian itself and the bulk translational mode, which is responsible for computing lengths of curves in the differential entropy formula \cite{holeo, Czech:2017zfq}. In the present case, the answer crucially involves non-local zero modes $Q^{(\cdot)}_{12}$ and $Q^{(\cdot)}_{21}$, which couple fermions in distinct intervals. Among them, we encounter a qualitatively novel zero mode $Q_{21}^{(2)}$, which is neither stress tensor-like nor a current.

It is also useful to contrast (\ref{curv1}-\ref{curv4}) with the holographic computation. At leading order in $1/N$, modular parallel transport describes deformations of RT surfaces. For two-interval regions in holographic CFT$_2$s, those surfaces can be in one of two phases: (i) one geodesic connecting $a_1$ with $b_1$ and another connecting $a_2$ with $b_2$, or (ii) the other way around---$a_1$ with $b_2$ and $a_2$ with $b_1$. Deformations of distinct geodesics commute, so non-vanishing curvature (at leading order in $1/N$) arises only for pairs that are connected by a geodesic in the given phase. The values of those non-vanishing curvature components are the same as in the single-interval case---because they describe the same deformations of a single geodesic. The single-geodesic curvatures were computed in \cite{mpt}. Borrowing those results, we find in phase (i) of a holographic CFT$_2$:
\begin{align}
[V_{a_1}, V_{b_1}] &\sim -\frac{2}{(a_1-b_1)^2} \int dz \frac{(b_1-z)(z-a_1)}{b_1-a_1}\, T(z)
\nonumber \\
[V_{a_1}, V_{b_2}] & \sim 0 \sim [V_{a_2}, V_{b_1}] \label{holocurv} \\
[V_{a_2}, V_{b_2}] & \sim -\frac{2}{(a_2 - b_2)^2} \int dz \frac{(b_2-z)(z-a_2)}{b_2-a_2}\, T(z) \nonumber
\end{align}
Symbols $\sim$ remind us that this result is subject to $1/N$ corrections. 

Expressions~(\ref{holocurv}) are of course given in $z$-space because there is no $X$-space for a holographic CFT$_2$. Despite that, we may think of them as loosely analogous to terms $Q_{11}^{(1)}$ and $Q_{22}^{(2)}$ in result (\ref{curv1}-\ref{curv4}). It is interesting to observe that terms, which in holographic theories populate two out of four curvature components, are distributed evenly among all four curvature components in the free fermion theory. This comparison can be interpreted as an imperfect toy model of $1/N$ corrections in holographic modular parallel transport. 

\section{Discussion} 
\label{sec:discussion}

We computed modular parallel transport in settings, which are novel in two aspects. To begin, ours is the first example of modular parallel transport of a disjoint region. It is also the first time where modular parallel transport is computable but not determined by holographic considerations.\footnote{The modular parallel transport of single intervals in all 1+1-dimensional conformal field theories is universal. Therefore, it is computable in non-holographic theories, but it takes a form which exactly mirrors that found in holographic theories.} The two novelties are interrelated: in holographic theories parallel transport of disjoint regions boils down to parallel-transporting their individual components; we were interested in the extent to which this holographic simplification fails in the free fermion. 

We found that non-local terms in modular parallel transport couple distinct components of a region for as long as it or its complement remain disconnected. The only limit in which modular parallel transport of two intervals $I_1 \cup I_2$ reduces to a local form is one where not only does $I_2$ shrink to zero length, but also the interval between $I_1$ and $I_2$ shrinks to zero---that is, when the problem literally boils down to the transport of a single interval. In particular, when $I_2$ was shrunk to zero size but maintained a finite distance from $I_1$, terms coupling the now point-like $I_2$ to $I_1$ continued to be important. In summary, the modular parallel transport of multiple intervals in the 1+1-dimensional free fermion theory is non-local whenever the kinematics allow it. 

\paragraph{A broader perspective} We would like to situate this finding in a broader context. In Ref.~\cite{Casini:2008wt,Agon:2021zvp}, the free fermion in 1+1 dimensions was found to be the unique conformal field theory whose mutual information in the vacuum is extensive $I(A:BC) = I(A:B) + I(A:C)$. This implies that, at the level of von Neumann entropies, the quantum entanglement in the 1+1-d free fermion vacuum is indistinguishable from that of a collection of EPR pairs \cite{withsirui}. Ref.~\cite{Agon:2021zvp} also gives an explicit formula for the mutual information of two arbitrary regions $I(A:B)$ in terms of conserved currents; this formula can consequently be interpreted as a density of entangled pairs. It would be interesting to understand the form of modular parallel transport in this language. Perhaps formulas~(\ref{dahmodz}-\ref{dbhmodz}) can be viewed as densities of entangled partners, which enter or leave a CFT region undergoing a deformation. 

Our method of calculation exploited the mapping between two intervals in the theory of one fermion and the half-line in a theory of two fermions. This mapping was key for computing the modular Hamiltonian, which was the starting point of our analysis. After that, the crisp form of modular parallel transport found in (\ref{dlambdaz}) was a consequence of the half-sided modular inclusion \cite{Wiesbrock:1992mg,Borchers:1999}. The relation between these facts and the extensivity of mutual information is not obvious. We hope to illuminate the link between these concepts in future work.

\paragraph{Multiple intervals} We have not attempted to compute explicit $z$-space or $X$-space expressions of modular parallel transport generators or curvature two-form for $(n>2)$-interval regions. They are unilluminating. Instead, we note that the obvious generalization of formula~(\ref{dlambdaz}) applies:
\begin{equation}
    V^{\text{($n$-int)}}_{\lambda} = \frac{1}{2\pi i} \sum^{n}_{i=1} \left[ - \frac{\partial a_i}{\partial \lambda} \partial_{a_i} \hmod + \frac{\partial b_i}{\partial \lambda} \partial_{b_i} \hmod  \right] \label{MPTN}
\end{equation}
This is because the relevant isomorphisms $\beta_n$ (see equation~\ref{betadef}) again imply that regions differing by $\pm da_i$ or $\pm db_i$ form half-sided modular inclusions. The non-localities, which we discussed in Sections~\ref{sec:limits} and \ref{sec:curvature}, will be qualitatively similar, that is bipartite. The ultimate reason for this appears to be the bipartite-like nature of the quantum entanglement in the free fermion vacuum \cite{Casini:2008wt}.

\paragraph{Comparison with the free boson}
Another theory, for which vacuum modular Hamiltonians of multiple intervals are known, is the free boson in 1+1 dimensions \cite{Arias:2018tmw}. Can we compute modular parallel transport there? This would be a difficult task. 

A schematic form of the two-interval modular Hamiltonian is:
\begin{equation}
\hmod = \int_{I_1 \cup I_2} dz \int_{I_1 \cup I_2}dw\, N(z,w)\, \partial \phi^\dagger(z)\, \partial \phi(w)
\label{coupling}
\end{equation}
Note that the coupling is not bipartite: the current $\partial \phi^\dagger(z)$ couples to currents all over both intervals. This non-locality is much more severe than in the free fermion. One way to understand it is that it reflects a GHZ-like character of quantum entanglement \cite{Casini:2008wt} in the free boson theory. Indeed, References~\cite{Casini:2008wt, Casini:2005rm} found that the entropy of a region $R$ in the 1+1-d massive free boson develops a term, which is independent of the number of connected components of $R$---a hallmark of GHZ-like entanglement. This term diverges in the massless limit. 

The greater degree of non-locality in (\ref{coupling}) appears to preclude the possibility of a modular inclusion. Without it, the only option for computing modular parallel transport is a direct calculation: (i) pick out zero modes from $\partial \hmod$, (ii) solve equation~(\ref{mptdef1}), then (iii) pick out zero modes again from the solution as demanded by (\ref{mptdef2}). The complicated nature of Hamiltonian~(\ref{coupling}) makes every step of this calculation quite intractable. The free fermion appears to be the unique theory, in which modular parallel transport of a disjoint region can be computed and understood. 

\section*{Acknowledgments} 
We thank Cesar Ag\'on, Pedro Martinez and Keyou Zeng for useful conversations. 
Both BCs and GW thank the organizers of workshop `Quantum Information and String Theory 2019' held at Yukawa Institute for Theoretical Physics, Kyoto University, where this work was initiated. 
The work of both BCs was supported by the Dushi Zhuanxiang Fellowship. 
LYH acknowledges the support of NSFC (Grant No. 11922502, 11875111) and the Shanghai Municipal Science and Technology Major Project (Shanghai Grant No.2019SHZDZX01).
GW is supported by the Center of Mathematical Sciences and Applications at Harvard University.

\end{document}